# Mining Interesting Trivia for Entities from Wikipedia

A DISSERTATION

*submitted towards the partial fulfillment of the*
*requirements for the award of the degree*
*of*

**INTEGRATED DUAL DEGREE**

*in*

**COMPUTER SCIENCE AND ENGINEERING**
**(With specialization in Information Technology)**

*Submitted by*
**ABHAY PRAKASH**

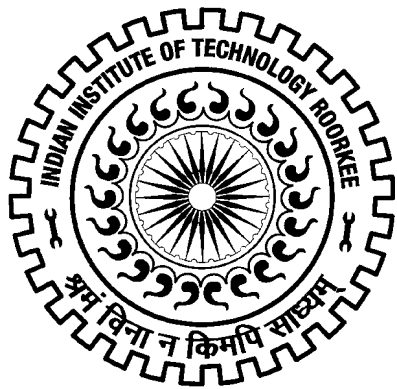

**DEPARTMENT OF COMPUTER SCIENCE AND ENGINEERING**
**INDIAN INSTITUTE OF TECHNOLOGY ROORKEE**
**ROORKEE – 247667 (INDIA)**
**May, 2015**

# CANDIDATE'S DECLARATION

I declare that the work presented in this dissertation with title "**Mining Interesting Trivia for Entities from Wikipedia**" towards the fulfilment of the requirement for the award of the degree of **Integrated Dual Degree** in **Computer Science & Engineering** submitted in the **Dept. of Computer Science & Engineering**, **Indian Institute of Technology, Roorkee**, India is an authentic record of my own work carried out during the period **from July 2014 to May 2015** under the supervision of **Dr. Dhaval Patel**, Assistant Professor, Dept. of CSE, IIT Roorkee and **Dr. Manoj K. Chinnakotla**, Applied Scientist, Microsoft, India.

The content of this dissertation has not been submitted by me for the award of any other degree of this or any other institute.

DATE : ........................    SIGNED: .........................................

PLACE: ........................                (ABHAY PRAKASH)

# CERTIFICATE

This is to certify that the statement made by the candidate is correct to the best of my knowledge and belief.

DATE : ........................    SIGNED: .........................................

                                    (DR. DHAVAL PATEL)
                                    ASSISTANT PROFESSOR
                                    DEPT. OF CSE, IIT ROORKEE

DATE : ........................    SIGNED: .........................................

                                    (DR. MANOJ K. CHINNAKOTLA)
                                    APPLIED SCIENTIST
                                    MICROSOFT INDIA (R&D) PVT. LTD.
                                    HYDERABAD, INDIA

## ACKNOWLEDGEMENTS

FIRST and foremost, I would like to express my sincere gratitude towards both of my guides **Dr. Dhaval Patel**, Assistant Professor, Computer Science and Engineering, IIT Roorkee and **Dr. Manoj K. Chinnakotla**, Applied Scientist, Microsoft, India for their ideal guidance throughout my entire graduate research period. I want to thank them for the advices, insightful discussions and constructive criticisms which certainly enhanced my knowledge and improved my skills. Their constant encouragement, support and motivation have always been key sources of strength for me to overcome all the difficult and struggling phases.

I thank **Puneet Garg**, Principle Development Lead, Microsoft, India for providing an opportunity to have the collaborated project between Microsoft, India and IIT Roorkee. His time-to-time discussions and valuable suggestions have been very important for the completion of the project.

I thank **Department of Computer Science and Engineering, IIT Roorkee** for providing lab and other resources for my graduate thesis work. I also thank **Microsoft, India** for providing the data required for the work.

I also thank **Sahisnu Mazumder** for being a selfless friend, for providing best suggestions and for always willing to help me in any problem both on and off water. Finally, I thank all my friends and lab mates for being very good friends, for a lot many good discussions and for being along with me in the whole journey and making it so joyful.

# ABSTRACT

TRIVIA is any fact about an entity, which is interesting due to any of the following characteristics – *unusualness, uniqueness, unexpectedness* or *weirdness*. Such interesting facts are provided in *Did You Know?* section at many places. Although trivia are facts of little importance to be known, but we have presented their usage in user engagement purpose. Such fun facts generally spark intrigue and draws user to engage more with the entity, thereby promoting repeated engagement. The thesis has cited some case studies, which show the significant impact of using trivia for *increasing user engagement* or *for wide publicity* of the product/service.

In this thesis, we propose a novel approach for mining entity trivia from their Wikipedia pages. Given an entity, our system extracts relevant sentences from its Wikipedia page and produces a list of sentences ranked based on their interestingness as trivia. At the heart of our system lies an interestingness ranker which learns the notion of interestingness, through a rich set of domain-independent *linguistic* and *entity based features*. Our ranking model is trained by leveraging existing user-generated trivia data available on the Web instead of creating new labeled data for movie domain. For other domains like sports, celebrities, countries etc. labeled data would have to be created as described in thesis. We evaluated our system on movies domain and celebrity domain, and observed that the system performs significantly better than the defined baselines. A thorough qualitative analysis of the results revealed that our engineered rich set of features indeed help in surfacing interesting trivia in the top ranks.

*I lovingly dedicate this thesis and all my achievements*
*to*

*my parents, my brother and my sister*

*for their endless love, support and encouragement.*

# TABLE OF CONTENTS

# LIST OF FIGURES

# 1

## INTRODUCTION

TRIVIA is any fact about an entity, which is interesting due to any of the following characteristics – *unusualness, uniqueness, unexpectedness* or *weirdness*. For example, for the movie *The Dark Knight (2008)*, a trivium could be - *"To prepare for Joker's role, Heath Ledger lived alone in a hotel room for a month, formulating the character's posture, voice, and personality"*. The sentence qualifies as a trivia as per our definition since it is unusual for an actor to seclude himself in a hotel for a month to just prepare for his role. These kind of facts draw the user to engage more with the entity since it appeals to their sense of appreciating novelty, curiosity and inquisitiveness, thereby promoting repeated engagement [1, 2]. A trivium could be presented either as a question-answer or as a single fact depending on the design of the experience and scenario.

## 1.1  Motivation

In the Internet age, where user attention span has become ephemeral, designing features and experiences which are not just usable but also *engaging*, has become the holy grail of all products and online applications. Failing to actively engage with the user may result in the user losing interest, getting distracted and finally abandoning or switching to a different application [3]. In view of this, researchers and practitioners have started designing product experiences which focus on the non-utilitarian aspect of the interaction which motivates the user to invest time, attention and emotion [2, 4]. For example, besides the search results, popular search engines surfaced rich experiences[1] such as interactive maps, polls and other statistics for election related queries during the U.S. Elections in 2014, and for cricket related queries during Cricket World Cup 2015 (CWC'15).

Figure 1.1: Example of Trivia consumption by Bing Search Engine [*ref.* Bing]

Figure 1.1 shows an example screenshot of result page surfaced for cricket related query on Bing during CWC'15. The rich experience composed of many statistical data, wining predictions, team ranking etc. which provide a lot of insights at one place. Such features increase the utility of the product or service. Along with the utility features, the result page also surfaced some interesting facts about cricket. In the example screenshot, the portion highlighted in red border presents a fact about a cricketer who excels in other games also like hockey, rugby, badminton etc. which is certainly quite interesting because of its unusalness. Such experience indulges the user to spend more time to browse through other presented trivia, and also lead him to return to the web search engine (Bing) in future.

---

[1] http://bit.ly/1njhlmh

## 1.2   Case Studies: Trivia used for user-engagement

Business cases studies [5–7] have presented some real world scenarios where trivia were used to *enhance user engagement* or for *wide publicity* of product/service as described below:

1. The first one presented in [5], is about a game – *Trivia Crack* [2] [3] in which users can submit trivia in question-answer form with multiple choices as options. Other users answer the trivia and then mark it as interesting or boring. Based on the votes, the boring trivia are continuously removed from the game. It has been the most popular app at Apple Store for continuous 66 days, which is the longest ever duration for any App being on top. To compare, the second longest streak as top app has been for 36 days by *Draw Something*, and has been only 5 days by the famous game *Candy Crush Saga*.

2. The second one presented in [6], is about Israel Democratic Institute using Trivia Game on their Facebook page for its promotion. A total of 6300 players played in the held two round of trivia game. Before the campaign, the page had 2700 likes, which increased to 7000 likes after first round and finally to 10,970 after the second round. This shows that the game was extremely successful for the wide publicity of target Facebook page.

3. The third one presented in [7] is about Voice Heard Media, which wanted to increase sponsor's email newsletter registrations. It embedded a Trivia product on company's website and Facebook tab. Users were free to take the first trivia question. But for the second one, they had to register. During registration process, an option to sign up for sponsor's email newsletter was given. Overall, 52.71% of unique visitors entered, and 33% of those opted in to sponsor's email news letter.

---

[2]App. on Google Play at http://bit.ly/16Jmb5f
[3]App. on Apple iTunes at http://apple.co/1RBeX6n

## 1.3   Proposal for Mining Trivia Automatically

The process of manual creation of trivia for any commercial purpose is both expensive and hard to scale across a large number of entities. In an experiment done within Microsoft, professional trivia curators were used for collecting trivia for some selected entities, and we observed that average throughput on one working day was around 50 trivia covering only 10 entities. Hence, in this thesis we introduce the problem of *automatically mining trivia for entities from unstructured text*. In Chapter 2, we have covered some approaches to generate trivia from structured databases, along with their limitations. The approaches have one or more of following limitations – not being fully automated, unavailability of data in required format and low variety in type of trivia. To explain what do we mean by "low variety in type of trivia", consider the example trivia for movie PK(2014) – *"Aamir Khan did not blink his eyes even once in complete movie"*. Such a trivia cannot be generated from structured databases, as per the data available in structured form.

Whereas, such limitations are not there in our proposed work as it has the following characteristics:

- **Fully Automated:** The approach discussed in this thesis is fully automated. The final results are usable as it is.

- **Ample Data:** Since the approach is on unstructured data, it means we can use any article, biography, news report etc. available over Internet. In this thesis, we has proposed Wikipedia as source of trivia, stating the reason in the following section 1.3.1.

- **Variety in Results:** The articles on Internet (Wikipedia for instance) contain a variety of information (facts) about the target entity in unstructured text form. In our experiments, we have obtained a variety of amusing facts as discussed in later chapters. A quick look on some actually mined trivia can be taken in APPENDIX-A

We propose to use Wikipedia to obtain candidate trivia for the entity, as explained in following subsection.

### 1.3.1 Wikipedia as a source of Trivia

*Wikipedia* is a community edited free encyclopedia. It contains pages on all kind of entities like movies, celebrities, countries, events etc. People who use it collaboratively create/edit the page for the desired entity, and provide information about it in natural language sentences, tabular or image form. Since it is community edited, a doubt may arise for correctness of facts due to accidental or deliberate malicious activity, know as *vandalism*. But, a detailed study presented in [8] states that "vandalism is usually repaired extremely quickly – so quickly that most users will never see its effects". Hence, Wikipedia can be treated as reliable for factual correctness.

We choose Wikipedia as our knowledge source since factual correctness is an important attribute for trivia, and Wikipedia provides sufficient number of trivia for most of the entities. To verify that this, we collected 100 trivia from web about a few selected entities and tried to find them in their Wikipedia page. We found that 56 trivia of those 100 trivia were present in entities' Wikipedia page, directly in a standalone sentence form. Sample of quality of trivia that can obtained from Wikipedia has been given in APPENDIX-D.

### 1.3.2 Approach Overview

We propose a novel approach called *"Wikipedia Trivia Miner (WTM)"* to mine interesting trivia about a given entity from its Wikipedia page. Since the application of trivia in user engagement requires few but really interesting facts, the thesis work targets to get the top-k most interesting trivia from the candidates. For the defined task, we propose to use an Machine-Learning (ML) approach. We tried and experimented with two ML methodologies – i) Classification methodology to identify whether a given sentence is trivia or not, and ii) Ranking methodology to rank the candidate sentences in order of their interestingness. The results obtained show that ranking methodology outperforms classification for obtaining top-k most interesting trivia. The comparison of the two methodologies have been presented in APPENDIX-C.

Being automated, the approach is scalable for any number of entities falling from any domain. We have experimented and demonstrate the effectiveness of WTM on two domains – movie entities and movie celebrities' domain to support our claim that *the approach developed is domain independent*. We used IMDB as a source for trivia for preparing the training dataset, as discussed in Section 3.1.1.

Although we chose Wikipedia as source because of aforementioned reasons, but one can always use other sources of trivia like books, news or any other archive providing facts about entities in natural language. Our features and approach are not limited to Wikipedia being source of trivia.

## 1.4 Problem Statement

Although, interestingness is a subjective notion which may differ from person to person, there are some facts for which there would be a significant agreement about their interestingness between a majority of people. In this work, we have currently restrict ourselves to such a majority based view of interestingness and leave the personalized subjective angle for future work. Hence, we define our problem statement as:

***Problem Statement:*** *For a given entity, mine top-k interesting trivia from its Wikipedia page, where a trivia is considered interesting if when it is shown to N persons, more than N/2 persons find it interesting.*

Discussion on statistically significant value of $N$, has been done ahead in 6.2.

Given an entity, WTM extracts relevant sentences from its Wikipedia page and orders them based on their interestingness using a machine-learning model. The final output of WTM is a list of top-k sentences which are *interesting trivia* for those entities.

## 1.5 Research Challenge

Since interestingness is a subjective notion which depends on cognitive aspects, it is extremely challenging to algorithmically capture the characteristics to identify whether a sentence is interesting or boring. e.g. Consider the trivia *"Aamir Khan did not blink his eyes even once in complete movie"*, from movie PK(2014). Most of the people will find it interesting because they have a cognitive knowledge that not-blinking is an usual thing. But performing the task algorithmically requires a lot of research effort.

# 1.6   Thesis Contributions

To summarize, the major contributions of our work are as following:

- *Introduced the problem* of automatically mining interesting trivia for entities from unstructured text source as Wikipedia.

- Proposed a system "Wikipedia Trivia Miner (WTM)" to mine top-k interesting trivia for any given entity based on their interestingness.

- Engineered features which can capture the about-ness (story) of the sentences and can help the ranker generalize which ones are interesting and which ones are not.

- Proposed a mechanism to prepare ground truth for test-set though crowd-sourcing. We experimentally found an optimum number of judges that should be used in crowd-sourcing, which is economic but simultaneously statistically significant to reflect the judgment of crowd.

# 1.7   Thesis Organization

Apart from the introduction and motivation in Chapter 1, rest of the report is organized as follows. Chapter 2 covers the literature review and presents our work in the context of related work. Chapter 3 presents an overview of system architecture. Chapter 4 covers the details for preparation of train dataset and feature extraction through it. Chapter 5 discusses theory of Ranker and mechanism of model building. Chapter 6 discusses the mechanism to retrieve top-k trivia from entity's Wikipedia page, and mechanism to evaluate WTM's performance. Chapter 7 discusses results both quantitatively and qualitatively and compares WTM's performance with considered baselines.

In Chapters 3,5 and 6, we have described the approach and demonstrated experiments only on *movie* entities to maintain the consistency for easier grasp. Chapter 8 covers the dataset description, experiments and results for *movie celebrities* to show the *domain independence* of the approach. Finally, Chapter 9 concludes the thesis and discusses the directions the work can be extended.

E ARLIER work in trivia mining focused on generating questions from structured databases. But, recently there is a noticeable trend in the text mining community to discover interesting items from unstructured text. In this chapter, we have covered works done in both the directions.

## 2.1  Related Work for Mining Trivia from Structured Databases

The work by Merzbacher et al. in [9] is the first attempt to automatically generate trivia from structured database. Figure 2.1, taken from [9] shows a sample database of Oscar awards, which has been used for experiments in their work.

| Year (date) | Category (string) | Title (string) | People (list of strings) | Win? (bool) | Cost (int) | Take (int) |
|---|---|---|---|---|---|---|
| 2001 | Best Picture | Gladiator | Wick, Frangoni, Lustig | T | | $187M |
| 2001 | Best Picture | Chocolat | Brown, Golden, Holleran etc. | F | $23M | $71M |

Figure 2.1: Example for structured Database [9]

They have given a functional approach to mine trivia questions in form of relational queries. The approach is to construct a series of functions on the relation, containing standard relational algebra operators. The relational queries thus obtained, are finally decoded by humans to frame natural language questions. However, the work is constrained to the availability of structured databases in the target domain, as well as is limited to discovering trivia which can be represented as structured queries such as:

$$\pi_{Film}(\sigma_{win=T}(\sigma_{year=2001}(\sigma_{Category=BestPicture}(Oscar))))$$

The above structured query represents the natural language query – "Which film won the Best Picture Oscar in 2001". For finding interesting trivia questions, all applicable pieces are scored. The high scored relational queries are taken as the most interesting candidates for further processing to get Natural Language questions.

The approach has some serious limitations. First, it is not completely automated as it requires human intervention for conversion of relational query to natural language query. Second, the paper states that the formulated queries start getting wild, complex, unreadable, and hard to be decoded by humans as the number of operators increases. Third, the weights for all possible operators are decided manually by a domain expert, based on the general observation of trivia questions formulated using those operators.

$$
\begin{aligned}
&\langle report\rangle \\
&\quad \langle date\rangle \; 23 \; april \; 2006 \; \langle/date\rangle \\
&\quad \langle location\rangle \\
&\quad\quad \langle city\rangle \; london \; \langle/city\rangle \\
&\quad\quad \langle country\rangle \; uk \; \langle/country\rangle \\
&\quad \langle/location\rangle \\
&\quad \langle event\rangle \\
&\quad\quad \langle action\rangle \; takeoverBid \; \langle/action\rangle \\
&\quad\quad \langle buyer\rangle \; megaCorp \; \langle/buyer\rangle \\
&\quad\quad \langle target\rangle minnowtech\langle/target\rangle \\
&\quad \langle/event\rangle \\
&\langle/report\rangle
\end{aligned}
$$

Figure 2.2: Example of structured news report [11]

The unusual, weird or surprising element of trivia could also be modeled using standard anomaly or outlier detection techniques. Author Byrne et al. in [10, 11] make such an effort in which they try to identify unexpected or surprising news by identifying the violation of expectation, formed using background knowledge and domain facts. As an input their approach takes structured news reports, similar to sample shown in Figure 2.2, taken from [11]. For the phenomenon captured in the the markup tags, they used to formulate an expectation. For instance, in the shown example, the structured report is about an event of an acquisition of one company by another. Generally, a bigger company takeovers a smaller company, and hence such news will be published frequently. So the expectation formulated will be that a bigger company acquires a smaller company. But if someday a smaller company takeovers a bigger company, then that will be a violation of expectation formulated and hence the news will be unexpected and surprising.

However, while mining trivia about an entity, background knowledge is not always obtainable and amenable to be modeled as expectation. e.g. consider the trivia *"The actors sang live on set"* (for movie *Jersey Boys (2014)*), which is an outlier and unexpected fact, but it is not feasible to model the notion that generally actors don't sing live on set. Moreover, the approach assumes that the input knowledge exists in structured format (XML) such as *<action> takeoverBid </action>*, as contained in Figure 2.2 also.Whereas, such rich structured data may not be available for all the facets of entities (such as actors singing live on sets).

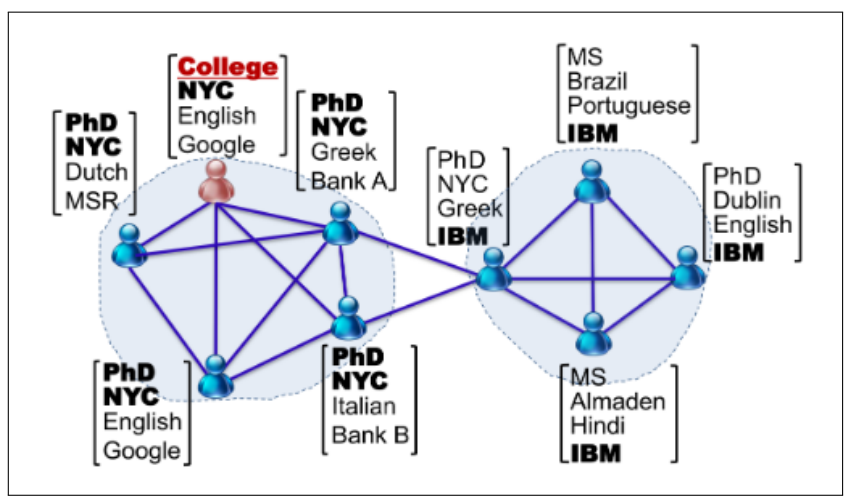

Figure 2.3: Example of attributed graph [12]

Authors Perozzi et al. in [12], have also given an approach to detect outliers in attributed graphs. Attributed Graphs are described where the objects(entities) are nodes, edges denote relations between them and each node has a feature vector associated with it. The component of feature vector represent various attributes, which the node can posses. Figure 2.3, taken from [12], shows an example of attributed graph. Note that each of the node has four attributes – highest degree, city, language and organization of working. By the given approach, first of all two clusters have been identified. Entities in same cluster are similar. Secondly, an outlier has been detected (one with missing PhD) in the left cluster for the reason that the every other member of this cluster posses a PhD. The authors have not utilized the approach directly for the purpose of mining trivia. But, entity trivia can be mined using the approach by providing it a similar attributed graph for entities. Knowledge bases like DBPedia [13], could have been used for the purpose.

But the approach is limited by the unavailability of attribute and attribute values for all the entities. For instance, for domain of Bollywood actresses with 782 entities, a set of 212 different attributes were found. Whereas for any given entity, the information was available hardly for 10 of these attributes i.e. the matrix was extremely sparse. For example, out of 782 entities, only 7 entities have the information for the attribute *Languages Known*. This doesn't mean that the other entities don't know any language. Also, as mentioned above, none of the structured database currently has complex facets of entities (such as actors singing live on sets).

The major limitation of such complex facets being unavailable in structured databases encouraged us to look in unstructured domain. Articles like Wikipedia page has ample amount of trivia present in them (as stated in Section 1.3.1). Hence, we chose unstructured text as the source for trivia where the above approaches can't be used. We investigated for the related work for identifying interesting things from unstructured text, as covered in the following section.

## 2.2 Related Work for Mining Trivia from Unstructured Text

Authors Gamon et al. in [14] presented a technique to identify interesting anchors from Wikipedia pages. They model the interestingness of an anchor by utilizing the users' browsing transitions within the Wikipedia domain, and formulate the problem as a click prediction task. However, their work is limited to anchors (links within the page). With the framed problem statement, the work reduces to click-prediction task to predict which anchors can be clicked while the user is browsing on current page. There is no natural language understanding involved to discover interesting non-anchor text. They used features like the position of the anchor, categories of the Wikipedia page, user's demographics like his geographic location, time of browsing etc. for learning purpose. Finally, they highlighted top $k$ anchors on the page, that reader will find interesting.

Authors Ganguly et al. in [15] have tried to identify aesthetically pleasing (beautiful) sentences, using only positive samples obtained from Kindle *"popular highlights"*. So, by the dataset it can be inferred that the model will predict whether a sentence will be highlighted by the reader or not. In particular, the authors engineered various features, which they used in their proposed one-class classification algorithm. Some of the prominent features were *Topic Diversity*, *Sentiment*, *Word Repetition* and *Part of Speech (POS)* tags. Whereas for trivia mining, except POS tags, other features are not relevant e.g., a trivia usually has only one topic, sentiment of a trivia may be positive, negative or neutral, and repetition of words in trivia sentences is not usual.

## 2.3   Research Gap

To the best of our knowledge, our dissertation work is the first one to propose a machine-learning based approach for mining trivia from unstructured text. Hence, there was a research gap at every step – dataset creation, source identification, designing approach and designing evaluation metrics to quantify the performance. Still, the major research contribution of the dissertation work has been in engineering domain-independent linguistic and entity based features which have been significantly successful to capture the subjective notion of Interestingness.

# SYSTEM ARCHITECTURE OF *Wikipedia Trivia Miner (WTM)*

I N Section 1.3, we surfaced an introductory overview of the approach to be used to mine interesting trivia from given entity's Wikipeida page. In this chapter, a detailed modular overview of architecture has been discussed. The chapter primarily describes source for each of the input and role of each module. The chapter also discusses the two execution phases of WTM. The detailed procedure of each module has been discussed in later chapters.

Figure 3.1 depicts the architecture of the entire system, in which our primary contribution lies in designing "Wikipedia Trivia Miner (WTM)". WTM is based on Machine-Learning approach to retrieve the most interesting trivia for the given entity. We have engineered a rich set of entity and language analysis based features to capture the signals of the sentence being interesting. With the extracted features, WTM's Interestingness Ranker builds a model over provided train dataset to learn to prefer sentences which could be interesting trivia over regular sentences.

WTM uses IMDB as source of trivia with which train dataset is curated. WTM uses entity's Wikipedia page as source to obtain candidates for trivia as discussed in Section 1.3.1. By leveraging the model built, the sentences in candidate set are ranked in order of their interestingness to obtain top-k most interesting trivia about the entity.

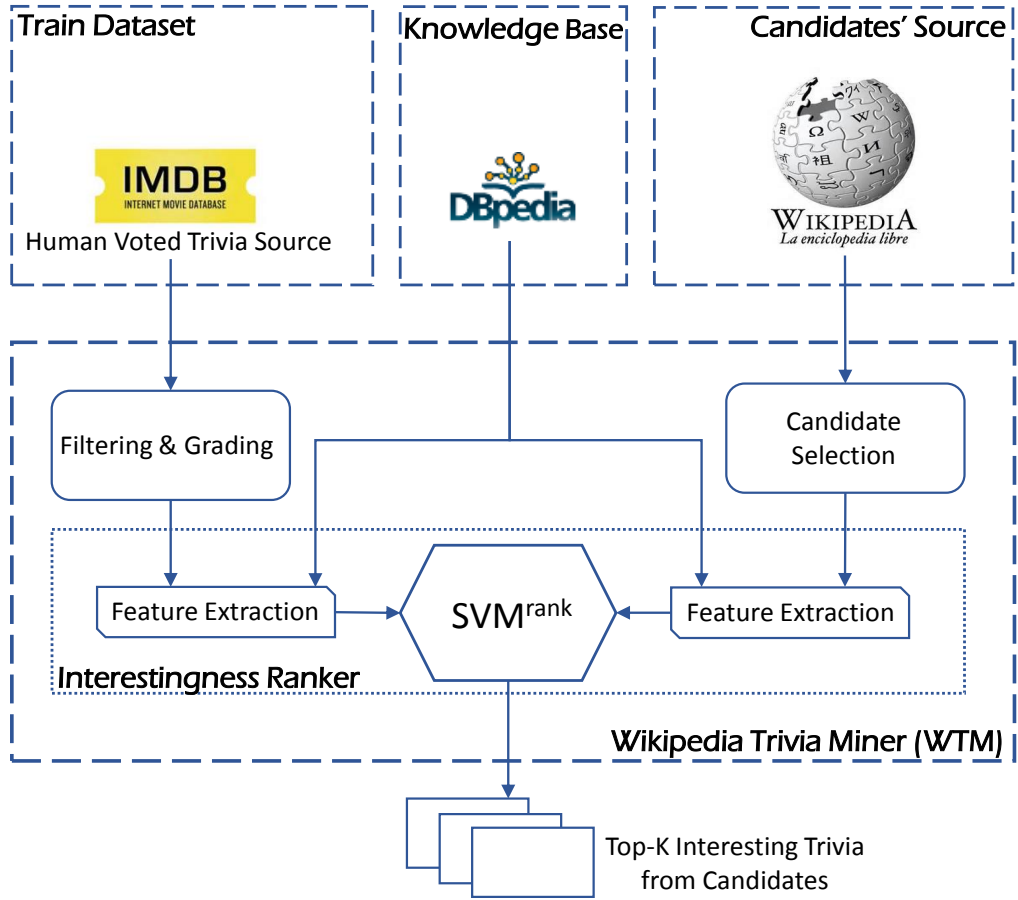

Figure 3.1: System Architecture of Wikipedia Trivia Miner (WTM)

The following sections describe in detail about the inputs required to WTM, various functional modules of WTM and the procedural phases of WTM system.

## 3.1 Input to WTM

For a given entity, WTM requires the inputs as described in following subsections.

### 3.1.1 Train Dataset

The approach requires training data (sample trivia) from the domain in which entity falls. Note that once training data is provided for a given domain, WTM can be used to mine trivia for any entity falling in that domain.

For movie entities, IMDB provides a dedicated page on trivia about the movie. Popular movies have usually more than 150 trivia. Figure 3.2 shows a screenshot of the page for movie *Batman Begins (2005)*. Note that there are 168 trivia for this movie. Moreover, each trivia is accompanied by the number of people voted for the trivia and number of people actually found it interesting. We crawled the respective page for the selected movies and thus created our raw set of sample trivia (along with the vote data).

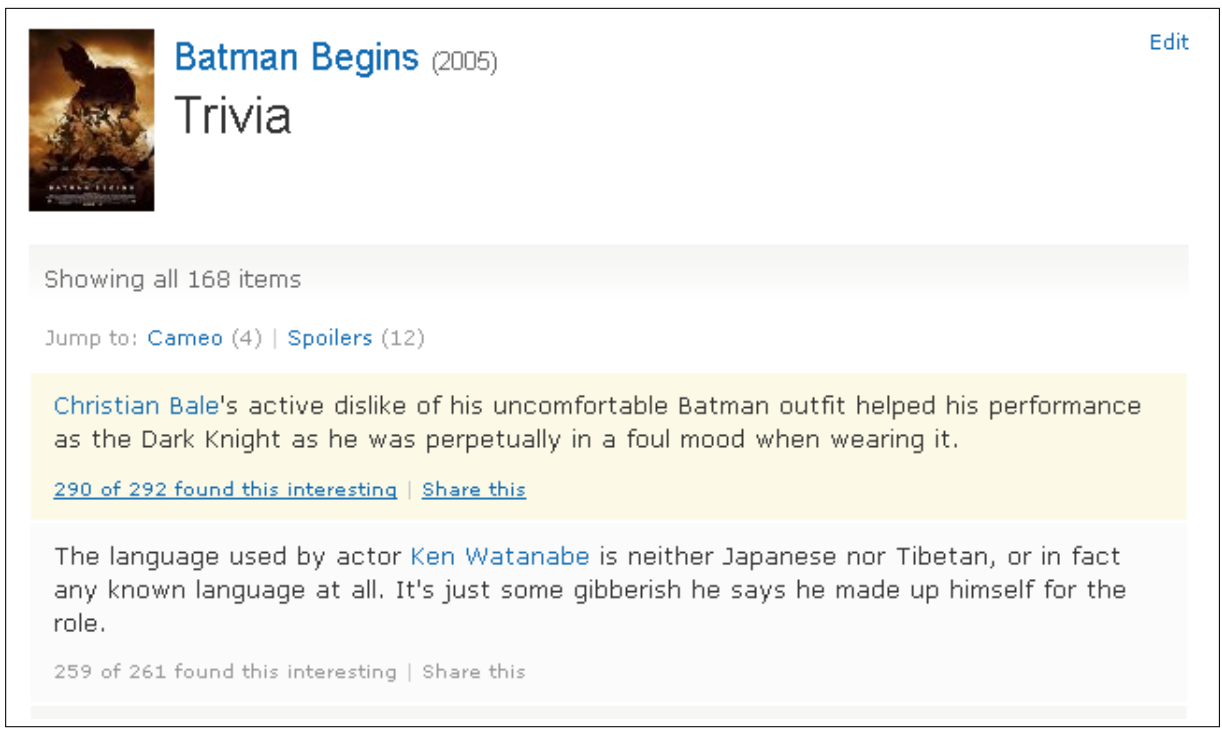

Figure 3.2: Snapshot of Trivia Page on IMDB for movie *Batman Begins (2005)* [*ref.* IMDB]

## 3.1.2   Candidates' Source

For a given entity, WTM needs some candidates (potentially trivia) to rank according to their order of interestingness. These candidates can be obtained from any article about the entity like – books, entity biography, news reports involving the entity etc. which contains grammatically correct factual sentences about the entity in natural language. Not all the sentences in the article are candidates as described in Section 3.2.2, and hence the module *Candidate Selection* extracts the candidate sentences for further ranking.

As discussed in Section 1.3.1, we have proposed to use only Wikipedia as candidates' source primarily because it is reliable of factual correctness [8]. Moreover, Wikipedia covers nearly all publicly available information about the entity. Hence, we can be assured of a large coverage for type of trivia. But, our features and approach are not limited to Wikipedia being source of trivia.

### 3.1.3 Knowledge Base

We have engineered some features which requires external knowledge, to know the relationship between the target entity and the mentioned entities in the sentence. A better clarity will arise once we describe our features in Section 4.3.3. We have proposed to use DBpedia [13] for obtaining the external knowledge.

For movie entity domain, we observed that IMDB (*Full Cast & Crew* page of the movie) provides much more number of relationships as compared to DBpedia. Hence, we crawled all the available relationships for the required entities from IMDB and formed our own knowledge base particular to the task at hand. Figure 3.3 shows a small screenshot, taken from the *Full Cast & Crew*[1] page of the movie *Batman Begins (2005)*. Note that in the screenshot, the information is present about who is the director, writer and who were the actors in the movie along with their played characters. The whole page contains several more information like producer, music director etc. of the movie.

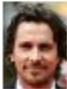

Figure 3.3: Sample relationships available on IMDB for movie *Batman Begins (2005)*[*ref.* IMDB]

---

[1] www.imdb.com/title/tt0372784/fullcredits

## 3.2   Modules in WTM

The WTM system has three primary modules: i) Filtering & Grading (FG), ii) Candidate Selection (CS) and iii) Interestingness Ranker (IR). Each of them have been described in following subsections:

### 3.2.1   Filtering & Grading (FG)

The data collected from IMDB (as described in Section 3.1.1) is consumed by Filtering & Grading module, which filters and labels the data into grades, based on their interestingness votes, as required to train our Interestingness Ranker. The procedural details have been discussed in Section 4.1.

### 3.2.2   Candidate Selection (CS)

As stated earlier, WTM uses entity's wikipedia page as source of trivia candidates. To prepare candidate set for trivia, the paragraph (<p>...</p>) tagged text is crawled from the Wikipedia page and passed on to Candidate Selection (CS) module. For any given movie entity, Wikipedia page covers different aspects of the target entity, using more than one adjoining sentences. But, not all of them could be understood independently without requiring appropriate context. For example, a sentence like *"It really reminds me of my childhood."* (from the movie *Let Me In (2010)*), is out of context and can't be understood independently. The Candidate Selection (CS) module selects sentences which are independently comprehensible. The procedural details have been discussed in Section 6.1.

### 3.2.3   Interestingness Ranker (IR)

Since for the task of user-engagement, one would need few but really interesting trivia. Hence, instead of testing each sentence for being trivia, we have proposed to obtain the most interesting top-k trivia about that entity. For such a task of ordering candidates, training a *Ranker* outperforms training a *Classifier* as discussed in Appendix-C along with the theoretical explanation. The appendix also presents the result metrics on final test set obtained by both the approaches, which clearly verifies the statement that *Ranker* is more appropriate choice than *Classifier* for the defined task. Hence, we go with the Ranking approach instead of classification approach.

## 3.3   Execution Phases in WTM

As standard in most of the Machine-Learning approaches, WTM too has two phases – *Train Phase* and *Retrieval (Test) Phase*. A model is built in the train phase using the samples labeled by humans. Then the built model is leveraged in the retrieval phase to obtain the ranking of candidates. Figure 3.4 shows the modules involved in two phases. Note that the model built in Train Phase is used in the Retrieval Phase.

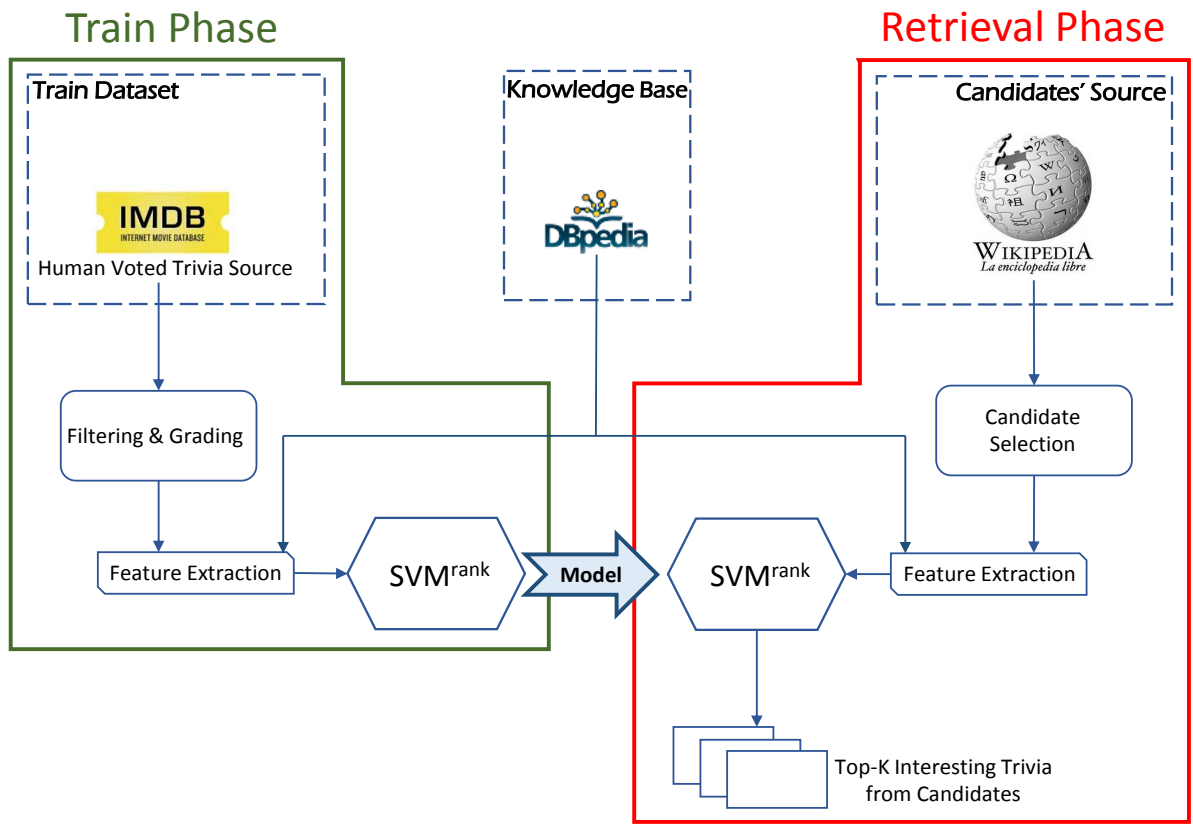

Figure 3.4: Modules in Two Phases of WTM

The particular tasks performed in the two phases are as following:

i) **Training phase:** This phase comprises of crawling of train dataset from IMDB, Pre-processing of train dataset to make it consumable by Interestingness Ranker through *Filtering & Grading* module, extracting features from the train samples through *Feature Extraction* module to map each of the sample in feature space and finally training *Ranker(SVM$^{rank}$)* in the feature space.

ii) **Retrieval (Test) Phase:** In this phase, the sentences from entity's Wikipedia page are crawled, potential trivia sentences are identified through *Candidate Selection* module, features are extracted from the selected candidates using *Feature Extraction* module again, to map them too in feature space and finally leveraging the already built ranking model to order the candidates according to their interestingness.

Chapters 4,5 discusses the details of Train Phase and Chapter 6 discusses the details of Retrieval Phase.

## MODEL BUILDING: PREPARING FEATURIZED TRAIN DATASET

I N Chapter 3, we gave an overview of various modules – covering *what* role each one of them plays. In this chapter, we discuss the details of procedure for the modules which are part of *Train Phase*. The chapter covers *how* to curate the train dataset through *Filtering & Grading* module, *features extracted* for training through *Feature Extraction* module.

## 4.1    Train Dataset Preparation

Section 3.1.1 describes about IMDB being the source of sample trivia. For movie domain, we selected 5000 most popular movie entities from Microsoft's internal Knowledge Graph, and crawled trivia for those from IMDB. In total we obtained 99185 such trivia. The IMDB trivia data also has a voting related interestingness measure in the form of *"X of Y found this interesting"*, where X is positive votes and Y is total votes for the trivia. Figure 4.1 shows an example screenshot of a trivia for the movie *Batman Begins (2005)*. Note that, the shown trivia has been voted by 90 people, out of which 86 have found it interesting.

Train data is crawled, along with the required meta-data (votes, movie name etc.) from IMDB source and passed to *Filtering & Grading (FG)* module, in which data is prepared for further consumption. Using the votes data, we calculate *Likeness Ratio (LR)* for each trivia as given by Eqn. 4.1

$$LikenessRatio\ (LR) = \frac{\#of\,Interesting\,Votes}{\#of\,Total\,Votes}$$
<div align="right">(4.1)</div>

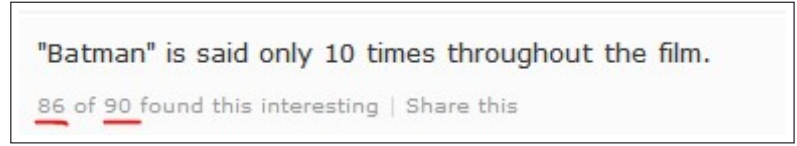

Figure 4.1: Sample Trivia for movie *Batman Begins (2005)* [*ref.* IMDB]

Since Likeness Ratio would be unreliable when computed only on a few total votes, we only consider trivia which have at least 5 total votes (minimum support). We observed that the distribution of trivia LR was highly skewed as shown in Figure 4.2 and followed a power law which is in agreement with earlier observations [16]. For instance, note that trivia with an LR of 1, were around 39.5% of the total.

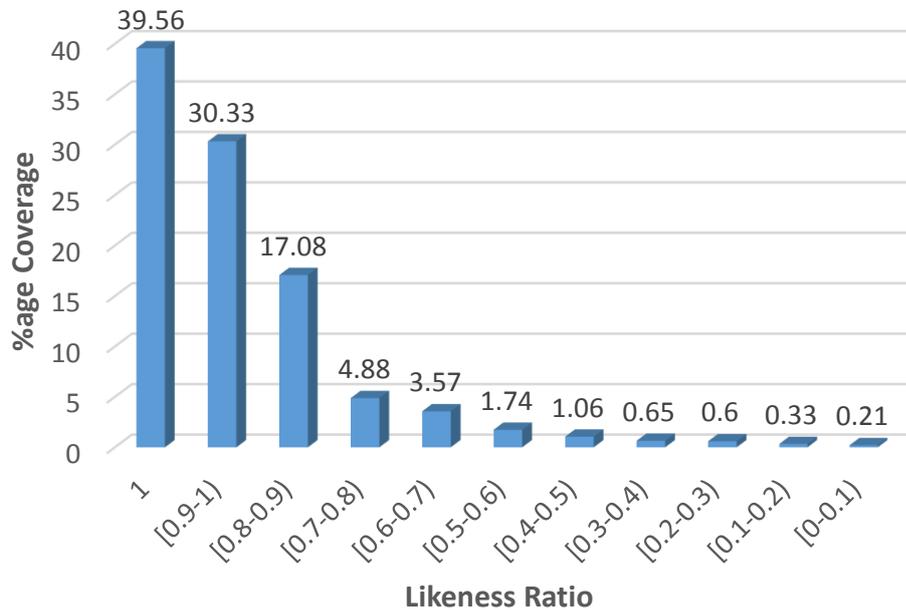

Figure 4.2: Distribution of Trivia over *Likeness Ratio (LR)*

Whereas for training a ranker, the distribution should follow a normal(Gaussian) distribution. Due to this, we discarded some of the trivia in higher L.R. range and so changed the minimum support only for higher LR ranges (greater than 0.6) to be 100 votes. We sort the remaining trivia based on their LR and assign grades to each of them by defining percentile cut-offs. The percentile cut-offs were placed at 90 (Very Interesting), 90-75 (Interesting), 75-25 (Ambiguous), 25-10 (Boring) and 10 (Very Boring). As a result of the above transformation, we ended up with 6163 trivia spanning across 846 movies, following the distribution as shown in Figure 4.3.

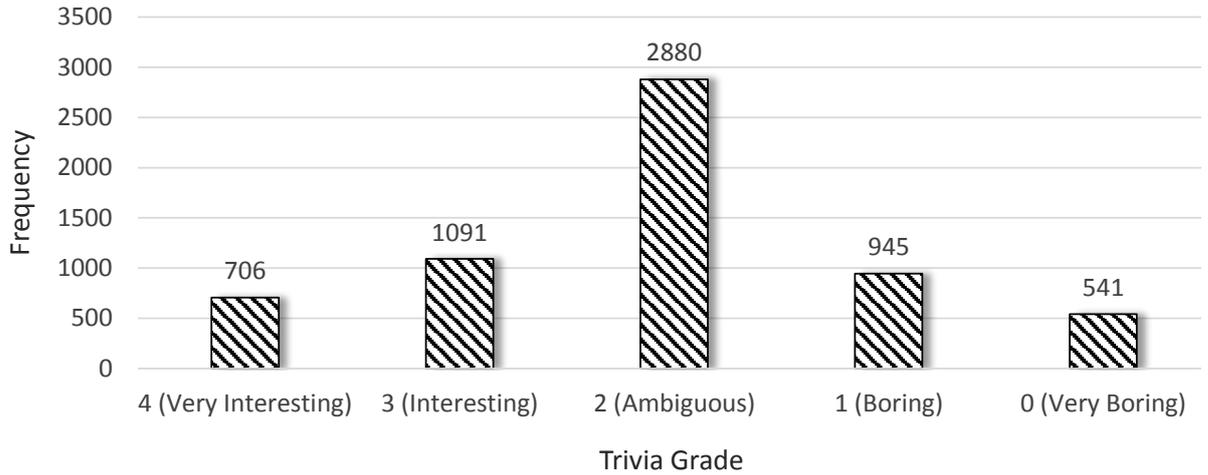

Figure 4.3: Frequency distribution of Trivia over defined grades

Table 4.1 gives the detailed statistics for each grade along with sample trivia. Note the examples and a marked difference in the degree of interestingness for each grade, which also gets reflected by the respective LR obtained.

Table 4.1: Sample trivia from each grade of IMDB data after the Filtering and Grading step.

| Grade | No. of Trivia | Sample Trivia | Movie Name | LR |
|---|---|---|---|---|
| 4 (Very Interesting) | 706 | Luc Besson wrote the original screenplay when he was in high school. | The Fifth Element (1997) | 1.00 |
| 3 (Interesting) | 1091 | Tom Cruise did all of his own stunt driving. | Jack Reacher (2012) | 0.98 |
| 2 (Ambiguous) | 2880 | Emily Blunt's character is named Rita, a possible nod to the love interest Rita from Groundhog Day (1993). | Edge of Tomorrow (2014) | 0.75 |
| 1 (Boring) | 945 | Andrea Riseborough was considered to play the role of Henley. | Now You See Me (2013) | 0.27 |
| 0 (Very Boring) | 541 | The first time Portia Doubleday and Rooney Mara are in the same movie since Youth in Revolt (2009). | Her (2013) | 0.20 |

## 4.2 Insight to ML being applicable for mining trivia

In the next section, we will be covering all the features that have been engineered for the task. But a question arise that *whether a sentence can provide characteristic signals of being interesting or boring by itself?* For answering this question, we present an observation here. In the crawled dataset, there were 32 such trivia which were repeated across the movies, and all the occurrences had nearly same likeness ratio.

As stated in Section 1.4, we treat a trivia as interesting if majority of people find it as interesting. This boils down to the conclusion that a trivia is *interesting* (Class 1) if $L.R > 0.5$ and it is *boring* (Class 0) if $L.R <= 0.5$. With this class definition, the respective repeated trivia because of having the same LR fell in the same class. This supports our belief that a sentence itself could consist some key characteristics for being interesting or boring. If we are able to engineer the correct set of features, the problem can be solved by Machine-Learning approach. **Table 4.2** shows few of the repeated samples. Note that even being presented for different movies, they have the nearly equal votes and LR, and hence fell into same class.

Table 4.2: Example Trivia repeated across movies.

| Movie | Trivia | Liked by | Total | LR | Class |
|-------|--------|----------|-------|-----|-------|
| American Beauty (1999) | The movie was named as one of "The 20 Most Overrated Movies Of All Time" by Premiere. | 48 | 124 | 0.39 | 0 |
| Good Will Hunting (1997) | --\|\|-- | 33 | 130 | 0.25 | 0 |
| Forrest Gump (1994) | --\|\|-- | 31 | 102 | 0.30 | 0 |
| Field of Dreams (1989) | --\|\|-- | 5 | 16 | 0.31 | 0 |
| The Dark Knight (2008) | In the early minutes of each film in the trilogy the main villain (Ra's Al Guhl, Joker, Bane) disguises himself as one of his own henchmen and there is a conversation about said villain in each scene. | 414 | 417 | 0.99 | 1 |
| The Dark Knight Rises (2012) | --\|\|-- | 325 | 329 | 0.99 | 1 |

## 4.3 Feature Extraction

The training input to IR is the graded interestingness data, in the form of (Movie, Trivia, Grade), as prepared in Section 4.1. The featurization step converts this triplet into a – (Movie, Features, Grade) where each trivia is transformed into a feature vector. The features extracted by the featurizer could be divided into three classes: Unigram, Linguistic and Entity based.

### 4.3.1 Unigram Features (U)

Using unigram features, we try to identify important words which make the trivia interesting e.g. words like *'improvise', 'award'* etc. might bring interestingness to the trivia for an entity from movie domain. For example in Table 2, the trivia from movie Jack Reacher (2012), is interesting because of the word *'stunt'*. We do some basic pre-processing before computing features: case conversion, stemming, stop word removal and punctuation symbol removal. We use TF-IDF weight of each unigram as feature.

### 4.3.2 Linguistic Features (L)

Mere unigram features are not enough to capture the semantics of sentences. Hence, we perform deeper language analysis on the trivia- POS tagging and dependency parsing to extract the following five different types of language-oriented features:

- *Superlative Words:* Words like first, best, longest, shortest etc. (of superlative degree) express the extremeness or uniqueness of the entity attribute being talked about and could be interesting. We use Stanford Core-NLP POS tagger to detect the presence of superlative adjectives (JJS) and superlative adverbs (RBS) in the sentence and fire a binary feature based on its presence.

- *Contradictory Words:* Presence of contradictory words indicates the presence of opposing ideas which could spark intrigue and interest. We borrowed a list of such words from online[1] source . Some of these words are *but, although, unlike* etc. We fire a binary feature based on its presence. An example trivia with such words is *"Although a very modest hit in theaters, it became one of the highest grossing video rentals of all time."*

[1] `http://bit.ly/1kOMshx`

- *Root Word of Sentence:* We use Stanford Dependency Parser [17] to obtain the complete parse tree of the sentence. The root word from a dependency parse helps in capturing the core activity being discussed in the sentence. e.g. *"Gravity grossed $274,092,705 in North America."* So, with gross as the root word, we can infer that the sentence is talking about some revenue related stuff. We mark the presence of root word in the form of a boolean feature *root_X*, where X is the lemmatized form of the word. Lemmatization means to bring the word in the root form of the word e.g. eat is the lemmatized form of ate.

- *Subject of Sentence:* We extract the subject of the sentence from the dependency parse as *subj_X*, where X is the lemmatized form of the word.

- *Readability Score:* Complex and lengthy trivia are hardly interesting. Hence, we use FOG Index as a feature − which is a measure of readability [18]. For a piece of text, FOG can be defined as given in Eqn. 2.

$$FOG = 0.4 * \left( \frac{\#ofWords}{\#ofSentences} + 100 * \left( \frac{\#ofComplexWords}{\#ofWords} \right) \right) \qquad (4.2)$$

We put the continuous score in one of the three bins $(0-7), [7-15]$ and $[15, \infty)$ based on the categories given by the same paper. Sentences with $FOG < 7$ can be easily read by school kids, FOG less than 15 can be read and understood by high school students and sentences with higher FOG are difficult to read and understand. Complex words are the words with more than two syllables.

### 4.3.3 Entity Features (E)

In order to learn entity and attribute-level generalizations in the model, we include named entities (using Stanford Named Entity Recognizer) and entity-linking features. For example, from the trivia *"De Niro was so anxious, he didn't attend Oscars"*, we would like to learn that - Entity.Actor not attending Oscars may be interesting. We include the following features:

- Presence of Generic NEs: Presence of NEs like MONEY, ORGANIZATION etc.

- Entity-Linking Features: We link NEs to entity attributes using knowledge from DBPedia knowledge base [13]. For example, if a trivia from movie *"The Fifth Element (1997)"* contains Named Entity *Luc Besson*, then it is linked to *entity_Director* as well as *entity_Writer*.

- Focus Named Entities of Sentence: For this feature, we resolve any NE present directly under the root. We mark its presence by a feature such as *underRoot_entity_Director*.

Note that, all the Entity Features are generic and not domain specific as while resolving the entities, we just look the `attribute:value` pairs for the entity from DBPedia. If any NE matches with the value in any `attribute:value` pair of entity, we replace the NE with *entity_Attribute*. For example, in case of country domain, while processing a sentence about the USA, all the occurrences of president's name, will be replaced by entity President. If the word is a NE, but still not resolved then, it is tagged as a feature *entity_unlinked_(NamedEntityType)* like *entity_unlinked_PERSON*.

To summarize, all our current features are generated *'automatically'* through: linking entity attributes using knowledge-base (DBPedia) and language analysis (Parsing, POS-tagging, NER etc.) on target sentence. In case of a new domain like Celebrities, the entity linking phase will automatically generate a different subset of domain-specific features such as entity_birthPlace, entity_Spouse etc. Based on the celebrity domain training data, the Rank-SVM model may assign higher weights to a different set of features than the ones in movie domain.

## MODEL BUILDING: TRAINING WTM'S RANKER

THE objective of the Interestingness Ranker (IR) is to rank candidate sentences in the decreasing order of their interestingness as trivia. We use the Rank-SVM [19] based formulation to automatically learn the ordering function from training data. This enables WTM to adapt to a new domain of entities by just changing the training data accordingly. In this chapter, we discuss the theory of Rank-SVM, give the actual reference of package used and give the parameters for which the best trained model is obtained. The chapter also discusses the contribution of each feature.

## 5.1 What is Rank-SVM?

To understand the concept of Rank-SVM, first we discuss the concept of *Support Vector Machines (SVM)* briefly.

### 5.1.1 Support Vector Machine (SVM)

Consider that there are two classes of training samples, and suppose we extract two features for each of the sample. Considering each feature as an independent axis, the samples can be represented by points placed in the formed two dimensional *feature space* based on the actual feature values they possess. For classification purpose SVM finds a hyperplane in the feature space, such that the plane is the bisector of maximum margin that can be formed between the two classes [20].

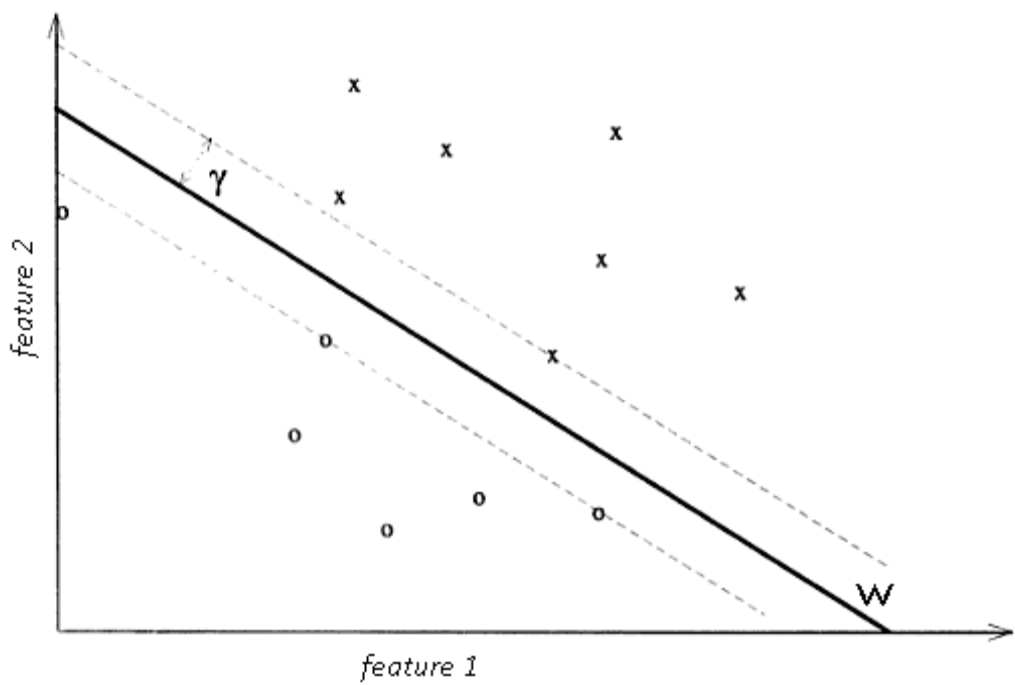

Figure 5.1: Finding hyperplane by margin maximization in SVM[20]

Figure 5.1, taken and modified from [20], explains the concept of SVM visually. Note that there are two classes which are represented as **o** and **x**. When the training samples have been mapped in defined feature space, the two classes are *linearly separable* which means a straight line (hyperplane in 2-dimensions) can separate the two classes. **W** represents the found hyperplane such that the margin between the two classes is maximum. The margin width is equals to $2 \times \gamma$.

For obtaining the class of an unseen sample, again the same set (ones which were extracted from training samples) of features are extracted for it, and the point is mapped to the defined feature space. Now, SVM predicts its class based on which side of the hyperplane the projection falls.

### 5.1.2 Rank-SVM

Unlike regular SVM, Rank-SVM tries to obtain a hyperplane in feature space, such that projections of each seen samples are in closest order as of their actual labeled order. For unseen samples, Rank-SVM order them according to the order of their projection on the found plane. Figure 5.2 demonstrates the training phase for a Rank-SVM. Figure 5.2 a) represents the format in which the input is given to the ranker. Note that for each sample, Movie_Id is also provided along with the labeled grade. Grade represents the relevance of the sample within the group. For our scenario, a trivia with higher grade will be more interesting for that movie as compared to another trivia with a lower grade hailing from the same movie. Reason for providing the Movie_Id is that Rank-SVM adopts *pairwise* approach, as explained in next paragraph. Figure 5.2 b) is just a visual representation for the process of model building.

Figure 5.2 c), taken and modified from Wikipedia, represents the model being prepared. Samples hailing from Movie_Id 1 have been shown in red colour and samples hailing from Movie_Id 2 have been shown in blue color. Note that the model has been prepared in form of hyperplane $W_1$, such that the projection of samples from Movie_Id 1 and Movie_Id 2 are simultaneously in the order of their actual grade, but within their own group. This is important to note that the sample with grade 2 from Movie_Id 1 is ahead of the sample with grade 3 from Movie_Id 2. Reason again lies in the concept of pairwise approach as described in next paragraph.

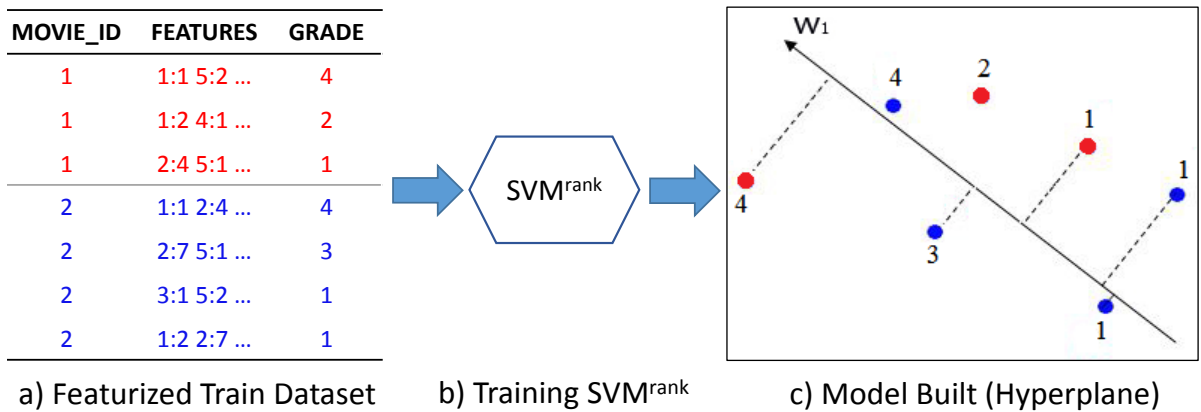

a) Featurized Train Dataset    b) Training SVM^rank    c) Model Built (Hyperplane)

Figure 5.2: Training of a Rank-SVM

$SVM^{rank}$ has been implemented over pairwise approach [21]. In pairwise approach, the model learns to harness available ordering within the group (here formed by Movie_Id). A pair of trivia (with different grades) is picked and the differences between two feature vectors are treated as new feature vectors. e.g. for three samples − $x_3$ with grade 3, $x_2$ with grade 2 and $x_1$ with grade 1, the new features will be Features($x_3$) − Features($x_2$), Features($x_3$) − Features($x_1$), and Features($x_3$) − Features($x_2$). So, for each pair ($x_3$,$x_2$) with rank($x_3$) > rank($x_2$), the pairwise approach additionally weighs the incremental features [Feature($x_3$) − Feature($x_2$)], hence capturing the specific signals which cause it to get higher rank. Note that for whole training, only samples within the formed groups are considered. Moreover, only the pairs of samples with different grade are considered.

Once the model has been prepared, ranking of unseen samples are done by preparing a similar featurized set for them. Unseen test samples are mapped in the feature space and their projection on the earlier found hyperplane $W_1$ are obtained. The order of projections is given by the order of *score* given as output for each sample. Hence, the samples can be sorted by the scores given to them for ranking them with in the group (again formed by Movie_Id). Figure 5.3 demonstrates the phase of ranking unseen samples. Note in Figure 5.3 a) that the input are similar to the one provided in train phase, but misses the grade. Figure 5.3 b) is just a visual representation of ranking procedure. Figure 5.3 c) shows the format of output obtained from the ranking procedure. Note that each of the sample has been given a real value score.

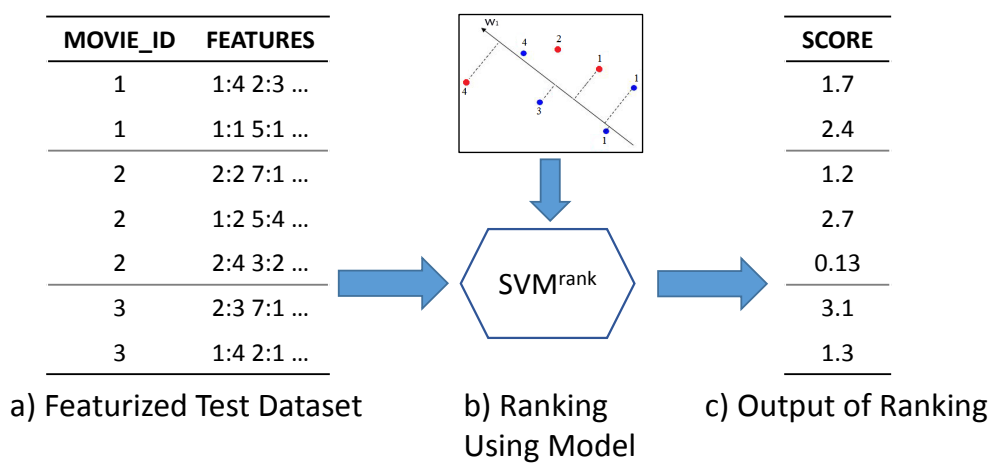

a) Featurized Test Dataset    b) Ranking Using Model    c) Output of Ranking

Figure 5.3: Ranking of Unseen Samples

## 5.2   Training Rank-SVM to obtain Best Model

We adopted the technique of 5 fold cross-validation (over the seen dataset), for engineering and selecting features and to build the model. We chose the model which gave the best results on cross-validate set. The *best model* refers to the model which was able to *generalize the most*. Finally, the chosen model is used to obtain results from unseen set (Wikipedia Page). To identify the best model, we need a metric on which the various models are evaluated.

### 5.2.1 Evaluation Metric for Intermediate Models

During training, the Rank-SVM model learns the feature weights and builds a model. The generalization capability of the model build depends on various parameters that can be set manually. The best model is identified by checking which model gives the best ranking performance at rank 10 for validation set. The ranking performance is measured using NDCG@10 [22], which is a standard metric used in IR to evaluate ranking.

### 5.2.2 Implementation used for Rank-SVM

We used the $SVM^{rank}$ package [19] for implementing the interestingness ranker. The resource page for obtaining the code or binary of $SVM^{rank}$ is given in APPENDIX-E.

We tuned kernel and two model parameters (by grid search) – $C$ (trade-off between training error and margin) and $e$ (epsilon: allow this much tolerance for termination criterion) using five-fold cross validation and rest of the values were set to default. The best parameters were found to be a linear kernel with $C = 17$ and $e = 0.21$. Figure 5.4 shows the plot for NDCG@10 obtained for different values of $C$ and $e$ for linear kernel.

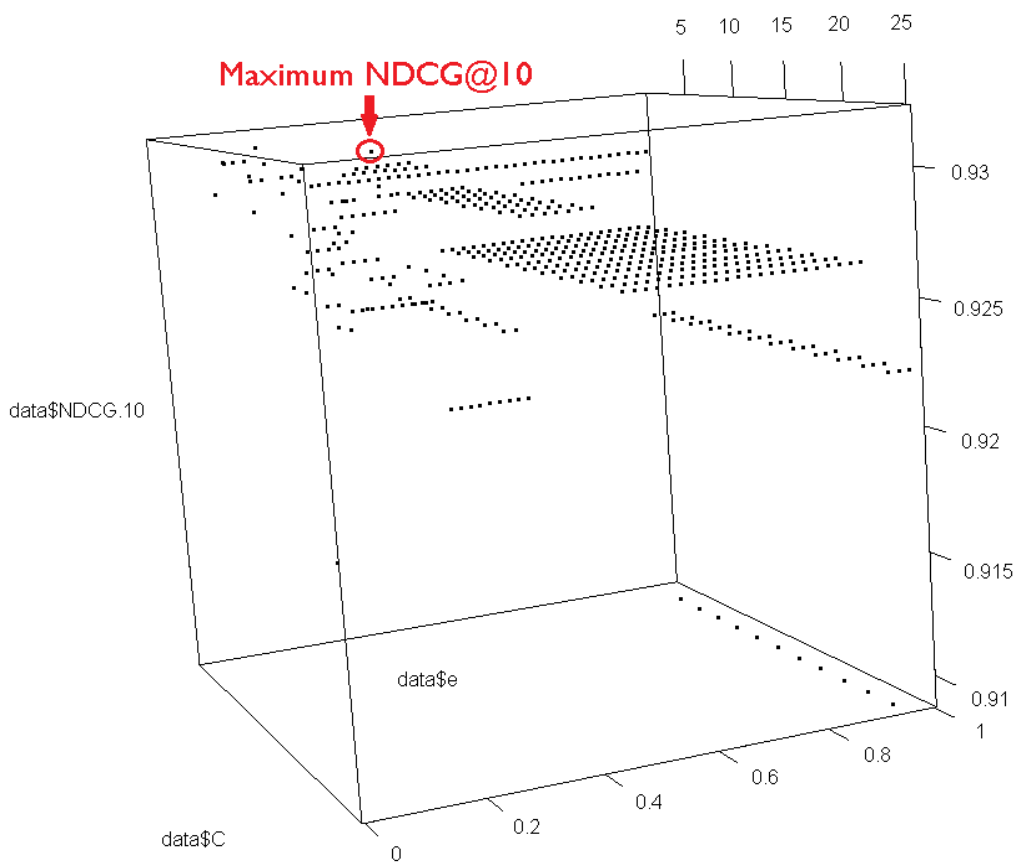

Figure 5.4: NDCG@10 varying with different $C$ and $e$

## 5.3   Feature Contribution

Table 5.1 shows the final results of the five-fold cross validation (training phase) with the respective best parameters of the model. The Entity Based Features such as Entity Attributes (Movie.Director, Movie.Producer, Movie.Actor), Named Entities (MONEY, ORGANIZATION etc.) result in the highest improvement followed by language analysis based features such as contradictions, root words and subject of dependency parse. The combined feature rich model outperforms all individual models.

Table 5.1: Results of five-fold cross validation with the best model parameters, with each feature group.

| Feature Type | No. of Features | NDCG@10 |
|---|---|---|
| Unigram (U) | 18025 | 0.934 |
| Linguistic (L) | 5 | 0.919 |
| Entity Features (E) | 4686 | 0.929 |
| U + L | 18029 | 0.942 |
| U + E | 22711 | 0.944 |
| WTM (U + L + E) | 22716 | **0.951** |

## 5.4   Generalization capability of engineered features

For any Machine-Learning task, the ability of predicting well for unseen samples is dependent on how much the model is able to generalize. In short ability to generalize provides the capability of *learning* ability to *predict* for unseen samples. Otherwise, the model will be able to give results (class/regression value/ranking) only for the seen samples (the ones which have been used to train the model), and that is as good as *memorization*. A well engineered set of features are important that can lead to better generalization. *A better generalization also gives the ability to learn more from less train data* as will be clear from the next paragraph, where we discuss the significant generalization provided by entity features.

In our engineered features, the entity features have played an important role for generalization. Table 5.2 shows top 20 most weighted features of WTM's interestingness ranking model. This gives a peek into the generalizations and domain knowledge related to trivia, learned by our model. Notice the features like *subj_entity_cast*, *entity_produced_by* etc. at higher ranks. Now suppose had we not done entity linking, then instead of the aforementioned features, we would have got features like *subj_batman*.

Now consider a scenario where train set has a trivia *"Christian Bale played the role of Batman"*. By entity linking done by us as described in Section 4.3.3, the model will get the sample as *"Entity.Actor played the role of Entity.Character"* and correlate the features with whatever grade the trivia falls in. Now, if the model is given unseen sample *"Aamir Khan played the role of P.K."*, the model will find it as a known sentence as the given unseen sample also reduces to *"Entity.Actor played the role of Entity.Character"*.

The higher weighted features also correlate with our general observations about movie trivia. For example, subject of a sentence being *scene*, main verb in the sentence is about *improvise*, words like *stunt*, *flop*, presence of entities such as Movie.Producer, Movie.Character etc. This shows how well the model has been able to generalize and identify which 'about-nesses' are found interesting by majority of people.

Table 5.2: Most informative features (by weight) from Interestingness Ranker Model along with their Feature Group

| Rank | Feature | Group | Weight |
|------|---------|-------|--------|
| 1 | subj_scene | Linguistic | 0.327141 |
| 2 | subj_entity_cast | Linguistic + Entity | 0.305082 |
| 3 | entity_produced_by | Entity | 0.225021 |
| 4 | underroot_unlinked_organization | Linguistic + Entity | 0.215818 |
| 5 | scene | Unigram | 0.214805 |
| 6 | root_improvise | Linguistic | 0.194474 |
| 7 | entity_character | Entity | 0.190772 |
| 8 | MONEY | Entity (NER) | 0.188747 |
| 9 | root_claim | Linguistic | 0.174173 |
| 10 | improvise | Unigram | 0.171668 |
| 11 | perform | Unigram | 0.165424 |
| 12 | root_reveal | Linguistic | 0.163567 |
| 13 | subj_who | Linguistic | 0.163417 |
| 14 | stunt | Unigram | 0.162217 |
| 15 | accord | Unigram | 0.161032 |
| 16 | superPOS | Linguistic | 0.158272 |
| 17 | subj_actor | Linguistic | 0.157705 |
| 18 | raccoon | Unigram | 0.156316 |
| 19 | real | Unigram | 0.155053 |
| 20 | root_change | Linguistic | 0.153183 |

## RETRIEVING TOP-k MOST INTERESTING TRIVIA

I N Chapters 4,5, we discussed how to obtain the build the model from training data. In this chapter, we discuss how to leverage the model to retrieve the most interesting trivia from the entity's Wikipedia page. We also discuss some metrics, on grounds of which we evaluate the performance of WTM. Needless to state that to quantify the performance, we first need to know the actual class of candidates. We discuss an efficient methodology to obtain the classes through crowd-sourcing. Finally, to compare our work we described some other methods also and chosen two appropriate ones out of them as baselines.

To retrieve the top-k most interesting trivia about the entity from its Wikipedia page, first of all Candidate Selection module is used to obtain the candidates from the Wikipedia text. Then the same features as described in Section 4.3 are extracted from each of the candidate. Finally the model built through training samples, is leveraged to order the candidates in order of their interstingness as trivia. The following section

## 6.1 Candidate Selection (CS)

As a first step in Candidate Selection (CS), we pre-process the Wikipedia page and extract the sentences present only in the paragraph environment. We ignore the text present in other environments such as infobox, tables, images, categories, references, links and itemized lists. We call the resultant text as *Core Content Text (CCT)*.

CCT, which is a set of sentences, also includes out of context sentences, as discussed in Section 3.2.2. To deal with this, there could be two solutions either provide context by including adjoining sentences or just drop such sentences for further processing. Guided by our intuition that shorter trivia are more readable than longer ones (multi-sentence), and hence are more likely to be interesting, we chose the later alternative of dropping out of context sentences.

Given the CCT of a target entity, we use Sentence Detector [23] to identify individual sentences. Next, we use Co-Reference Resolution, to find out links between sentences in a given paragraph, and remove those sentences which have mentions outside the current sentence. However, sentences which refer to the target entity are retained. For instance, for the movie *Forrest Gump (1994)*, the following are sample sentences from Wikipedia *"Hanks revealed that he signed onto the film after an hour and a half of reading the script. He initially wanted to ease Forrest's pronounced Southern accent, but was eventually persuaded by director Bob Zemeckis to portray the heavy accent stressed in the novel."* The first sentence has an outlink as *the film* which refers to the target entity. The second sentence refers to the hero "Tom Hanks" as *He*. As stated earlier, in CS phase, we drop the second sentence and retain the first one.

### 6.1.1 Limitations of approach used for CS

Note that few of the dropped sentences could have been retained by replacing the pronoun with the actual entity using co-reference resolution. But, we right now refrained from the technique and targeted only self-contained sentences which are independently comprehensible. Resolving co-references was not used because it may still lead to incomprehensible trivia e.g. replacing 'I' in "I don't know." will still not make it self-contained and comprehensible. Recall can be certainly be increased by proper substitution, and we propose the task as future work.

Due to dropping the out-of-context sentences, WTM must be missing some of the interesting trivia which span over more than one sentence. To retain such kind of trivia, we need to apply some technique which can identify which of the sentences (minimum number) are related and become complete in themselves. Such a task is itself a research challenge, and we propose the task as a future work.

### 6.1.2   Effect of Candidate Selection

Since our original problem focuses to obtain top-k interesting trivia, we quantify the performance of Candidate Selection module with its contribution to enhance the final quality. Table 6.1 shows the effect of adding CS module on WTM final accuracy for both uni-gram and the final WTM models. CS improves the precision of the final WTM model by more than 16%. These results prove that CS is indeed helping in eliminating out of context sentences.

Table 6.1: Effect of CS on WTM Precision. WTM (U) is only Unigram and WTM (U+L+E) is final system.

| Model | P@10 before CS | P@10 after CS | % Improv. |
|---|---|---|---|
| WTM (U) | 0.28 | 0.34 | 21.43 |
| WTM (U+L+E) | 0.39 | 0.45 | 16.67 |

CS module had a recall of 0.49 over all the sentences judged from set of Wikipedia sentences. But, although human annotators marked them as interesting, most of the dropped sentences required context of previous sentences and paragraph. For an instance for the movie *Pineapple Express (2008)*, its Wikipedia page contained a paragraph on a leakage incident of movie's red-band trailer. The paragraph contains a sentence which is a statement given on this incident by a related person – *"Interscope asked me and I was, like, well, since it's just the trailer, that's cool. I didn't really think twice about it"*. The sentence detection module isolated the sentence *"Interscope asked me and I was, like, well, since it's just the trailer, that's cool."*, and it was sent for judgement. Human judges might be knowing the incident beforehand, and hence marked it as interesting(5 judges out of 5). Undoubtedly, the overall scenario is interesting, but the standalone sentence is not comprehensible without paragraph context. Hence, we justify our method of evaluating the module's performance only by its final contribution towards increase in final accuracy.

## 6.2   Test Dataset Creation to evaluate WTM

In order to evaluate the effectiveness of our system, we created a dataset of sentences from the Wikipedia pages of movie entities. As shown in Table 6.2, we downloaded the Wikipedia pages for 20 movies randomly sampled from the top 5000 popular movies of IMDB. As described in Section 6.1, we pre-processed and extracted the CCT for all the movies. Later, we divided them into sentences and obtained crowd-source judgments on their interestingness. In order to avoid subjective bias, each movie, sentence pair was judged by five judges on a two-point scale (Interesting, Boring). The judges were given detailed guidelines along with sample judgments. As per our defined majority based notion of interestingness, the majority judgment was then marked as the final label for the trivia. In our entire experiments, to be fair, the input to all the approaches (Baselines and WTM) was given as preprocessed Wikipedia text (CCT).

Table 6.2: Dataset Details

| Dataset Name | Source | No. of Sent. | No. of Movies | No. of Pos. Sent. (Trivia) |
|---|---|---|---|---|
| Train Set | IMDB | 6163 | 846 | - |
| Test Set | Wikipedia | 2928 | 20 | 791 |

Since interestingness is a subjective notion, getting as many judgments as possible would be ideal for making reliable conclusion. However, we chose the number of judges as five due to following reasons. Each Wikipedia page has 100 sentences on an average. So, due to practical budgetary limitations, we could not go beyond the current number of judgments.

However, we performed an experiment to validate the extent to which our crowd-sourced judgments with five judgments per trivia match the general wisdom of the crowd. Since the IMDB train data has many votes for each trivia, we randomly sampled 100 trivia from it and as per our defined notion of interestingness, we labeled those with Likeliness Ratio (LR) greater than 0.5 as interesting (class 1) and LR less than 0.5 as boring (class 0). This is essentially a majority vote on the total votes polled. We call these labels as IMDB vote based labels. Later, we got 5 judgments for each of these 100 trivia through our crowd-sourcing platform and assigned labels based on majority vote. We call these labels as Crowd-Source labels. We calculated the agreement between these two labeling mechanisms (IMDB vote based labels and Crowd-Source labels) using the Kappa Statistic [24]. We found the Kappa value to be 0.618 which means the agreement is *substantial* (Refer APPENDIX-B for more details on Kappa Statistic). This shows that five judges although not ideal, are sufficient to adequately reflect the general wisdom of the crowd.

## 6.3   Evaluation Metrics for Final Results

Given a Wikipedia page for an entity, our system produces a ranked list of 'top-k' most interesting trivia sentences. Hence, we use *Precision@k* as our evaluation metric. This means that out of top-k results, how many actually are interesting trivia.

However, to demonstrate the effectiveness of our system in terms of its ability to bring in diverse kinds of trivia, we also report *Recall@k*. This means that out of all trivia present in candidate set, what fraction has already been retrieved till rank k. This graph will be monotonously increasing, and quality of ranking is measured by how steeply the graph rises.

## 6.4   Comparative Approaches and Considered Baselines

We have compared our WTM (with advanced features) with six other approaches. We define two approaches among them as baselines, against which we will compare the performance of our system. The defined approaches are as following:

- **Random Pick**: The first mechanism is a *random* (Baseline I), in which we randomly picked any 10 sentence from the Wikipedia page (only CCT) of the entity. Since there did not exist any prior approach for the task, we chose this first mechanism as our first baseline.

- **Random Pick after CS**: Our second mechanism is randomly pick sentence, but after removing the out-of-context sentences i.e., set obtained after Candidate Selection module. Note that, this mechanism signifies randomly picking sentences from standalone meaningful sentences.

The next mechanisms are in which the obtained candidates were ranked by applying different ranking algorithms. Note that superlative words such as best, most, highest, largest are often used to show the uniqueness, which makes sentences interesting. For example, for the movie *The Matrix (1999)*, the following sentence is a trivia – *"In 2007, Entertainment Weekly called The Matrix the best science-fiction piece of media for the past 25 years."*

Hence, the first of the three ranking approaches are where the candidates have been ranked by the number of superlative words (sup. POS) present in them. We use the Stanford POS Tagger for identifying these superlative words. Superlative words have their Part of Speech either as superlative adjective (JJS) or superlative adverb (RBS). Note that since not all the sentences with superlative word(s) are interesting, there could be a situation when more than 10 such sentences containing a superlative word are present, but not all are interesting. In such a scenario, the metrics will depend on the permutation got selected to be in top 10. Hence, we have defined three mechanism of deciding the permutation as following:

- **CS + Ranking by Sup. Words (Worst Case)**: Here we deliberately chose the permutation in which for the same number of superlative words, un-interesting sentences (with label 0) were placed higher. This metric would give a consumer an idea that if he uses this algorithm, what worst would he could get.

- **CS + Ranking by Sup. Words (Random Case)**: We shuffled the sentences and then sorted them based on the number of superlative words present in them. Such an experiment was done 5 times, to report the average metrics.

- **CS + Ranking by Sup. Words (Best Case)**: Here we deliberately chose the permutation in which for the same number of superlative words, interesting sentences (with label 1) where placed higher and hence we would get them in top 10 before the uninteresting ones. By using the approach of ranking by number of superlative words, one would never obtain metrics better than the Best Case. We chose this mechanism as Baseline II, so as to take a stronger baseline to compare with.

The last two mechanisms are the ones where WTM was used for ranking, over the set of candidate sentences obtained by Candidate Selection. The approaches are as following:

- **CS + Ranking by WTM (U)**: In this mechanism, we used $SVM^{rank}$ trained using Unigram features only.

- **CS + Ranking by WTM (U+L+E)**: In this mechanism we used $SVM^{rank}$ trained using all the engineered features. This mechanism signifies our complete approach.

## RESULTS AND DISCUSSION

THE previous chapter discussed how to retrieve top-k most interesting trivia about the entity from its Wikipedia page. The chapter also discussed the overall mechanism to quantitatively evaluate WTM in comparison to other define approaches and chosen baselines. In this chapter, we discuss the results obtained on the defined metrics for WTM, and compare them with the baselines. The chapter also presents a qualitative analysis identifying the scenarios where WTM outperforms other approaches, and also the areas where WTM lacks behind other approaches. Finally the chapter discusses the effect of train dataset size on performance of WTM.

## 7.1 Quantitative Result

WTM has been quantitatively compared with other baselines on the two metrics, as discussed in following subsections.

### 7.1.1 Evaluation by Precision at 10 (P@10)

As stated earlier at many places, our defined task is to retrieve top-k most interesting trivia from entity's Wikipedia page. We compare all the approaches taking standard value of $k = 10$. Table 7.1 shows the overall results of our system in comparison with the other approaches. Note that in Precision@10 metric, WTM system performs significantly better than both the baselines with an improvement of 78.43% over Random and 33.82% over Superlative POS (Best) techniques. Note that WTM (U+L+E) Features has also achieved an improvement of 33.82% over the plain WTM with (U) features. We also did statistical significance test while comparing with baseline approaches. Results marked as ‡ indicate that improvement was statistically significant at 95% confidence level ($\alpha = 0.05$) when tested using a paired two-tailed t-test.

Table 7.1: Results comparing the performance of WTM with other approaches.

| Model | | Avg. P@10 | % Improv. (Baseline I) | % Improv. (Baseline II) |
|---|---|---|---|---|
| Random Pick from Wikipedia Page (**Baseline I**) | | 0.25 | - | - |
| After Cand. Selection (CS) | Random Picked | 0.30 | **19.61** | - |
| | # of Sup. POS (Worst Case) | 0.32 | **27.45** | - |
| | # of Sup. POS (Random) | 0.33 | **29.41** | - |
| After Cand. Selection (CS) + Ranking By | # of Sup. POS (Best Case) (**Baseline II**) | 0.34 | **33.33**‡ | - |
| | WTM with (U) Features | 0.34 | **33.33** | 0 |
| | WTM with (U+L+E) Features | 0.45 | **78.43**‡ | **33.82**‡ |

## 7.1.2 Evaluation by %age Recall at k

Figure 7.1 compares the recall of WTM with other baselines. Note that, although the Superlative POS fares well initially vis-a-vis recall, it reaches saturation very soon. This is due to the fact that Superlative POS baseline lacks diversity and can only retrieve a single type of trivia - those which contain superlative descriptions. On the other hand, besides superlatives, WTM can retrieve a variety of other trivia due to its rich feature based on language analysis and entity understanding.

The absolute percentage recall shows the qualitative performance too. Note that, % recall till rank 3 is more for SuperPOS approach as compared to WTM. This signifies that, had we required only 3 or lesser number of trivia, use of SuperPOS approach would give a better performance i.e. there are more chances that SuperPOS gets more number of interesting trivia in three results as compared to WTM. Whereas, WTM outperforms SuperPOS approach soon after rank 3, which means that if we need and mine more than 3 trivia, WTM will obtain more number of interesting trivia as compared to SuperPOS approach. Also note that at rank 25, WTM has an average recall of around 60%. This is indeed a valuable proposition which enables mining of entity trivia at scale.

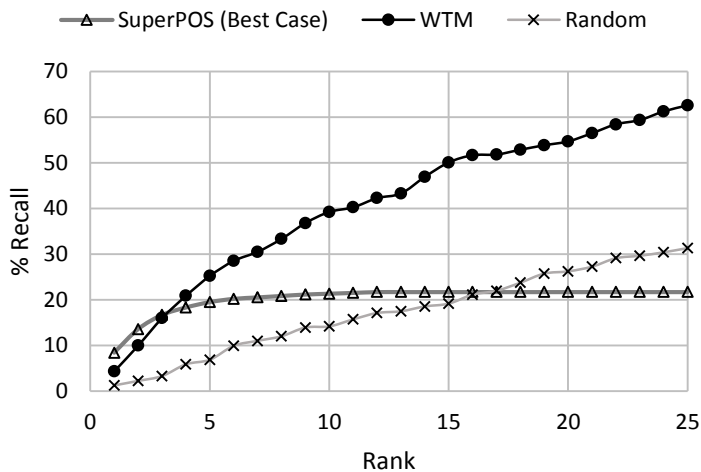

Figure 7.1: Comparison of Recall at various ranks of WTM with baselines on Wikipedia Test Set.

## 7.2 Qualitative Comparison

Table 7.2 shows the qualitative analysis of trivia mined from WTM and Sup. POS approaches. The table presents representative samples of both interesting and non-interesting trivia mined along with an explanation. The first section, WTM Wins (Sup. POS Misses), represents those samples where WTM retrieved the interesting trivia in top-10 and the baseline Sup. POS completely missed it. In the second example of this section, from *Gravity (2013)*, WTM gets the trivia correctly due to – a) resolving the NE *Cuarón* to Entity.Director b) getting the subject of the sentence as the *script*, c) important unigram features such as *film, years* and finally d) the ranker understanding that this combination of features is interesting. Sup. POS doesn't have any such sophisticated features and hence totally misses it.

The second section of table, WTM's Bad, shows some weaknesses of the model due to which non-interesting trivia show up. The first example shows a failure of CS that allows an out of context sentence. The second example shows a case where the trivia was ranked important due to the root verb *receive* getting undue importance.

Table 7.2: Qualitative comparison of trivia mined using WTM and Sup. POS techniques.

| Result | Movie | Trivia | Description |
|--------|-------|--------|-------------|
| WTM Wins (Sup. POS Misses) | Interstellar (2014) | Paramount is providing a virtual reality walkthrough of the Endurance spacecraft using Oculus Rift technology. | Due to Organization being subject, and (U) features (technology, reality, virtual) |
| | Gravity (2013) | When the script was finalized, Cuarón assumed it would take about a year to complete the film, but it took four and a half years. | Due to Entity.Director, Subject (the script), Root word (assume) and (U) features (film, years) |
| | The Deer Hunter (1978) | De Niro was so anxious that he did not attend the Oscars ceremony. | WTM gets it right due to Entity.Actor, Root word(anxious) and (U) features (oscars, ceremony) |
| WTM's Bad | Elf (2003) | Stop motion animation was also used. | Candidate Selection failed |
| | Rio 2 (2014) | Rio 2 received mixed reviews from critics. | Root verb "receive" has high weightage in model |
| Sup. POS Wins (WTM misses) | The Incredibles (2004) | Humans are widely considered to be the most difficult thing to execute in animation. | Presence of 'most', absence of any Entity, vague Root word (consider) |
| Sup. POS's Bad | Lone Survivor (2013) | Most critics praised Berg's direction, as well as the acting, story, visuals and battle sequences. | Here 'most' is not to show degree but instead to show genericity. |

There are some cases where the baseline Sup. POS got some good trivia which WTM missed (top-10). The third section presents an example from *The Incredibles (2004)* which consists of a superlative word *most*. However, we couldn't get it due to absence of any prominent entity (no entities), language (vague root word: consider) or unigram related features.

The fourth section presents an example from *Lone Survivor (2013)* where Sup. POS fails and retrieves a non-interesting trivia. This is due to miss-classifying the word 'Most' occurring at the beginning of the sentence as superlative. For more samples of actual result obtained by WTM, refer APPENDIX-A.

## 7.3   Sensitivity to Training Size

As given in Table 6.2, in our current system, we use around 6K trivia training samples filtered from around 846 movies. However, we also studied the effect of varying training data size on the precision of our system and report it in Figure 7.2. Results show that the precision of WTM increases with the size of training data. This is a desirable property as it allows us to further improve the precision of WTM by including more training data (for instance, by expanding the initial movies list beyond 5K)

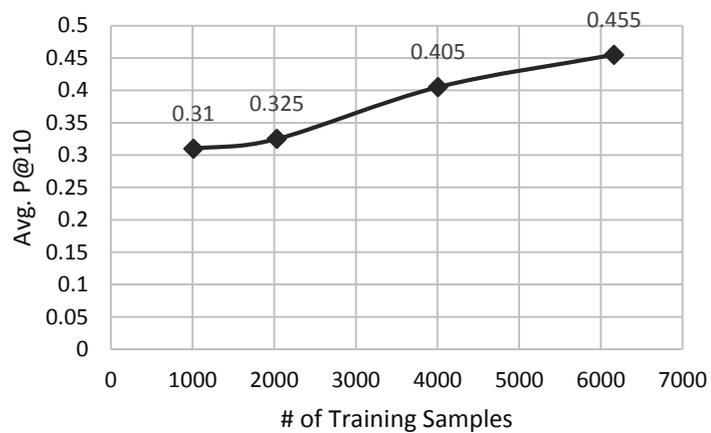

Figure 7.2: Training Data Size vs. WTM Accuracy

# WTM'S DOMAIN INDEPENDENCE: EXPERIMENT ON MOVIE CELEBRITY DOMAIN

AS claimed and described earlier also in Section 4.3.3, our engineered features are generated *'automatically'* through: linking entity attributes using knowledge-base (DBPedia) and language analysis (Parsing, POS-tagging, NER etc.) on target sentence. While in movie domain the automatically curated entity features were like entity_producer, entity_director etc. in case of a new domain like Celebrities, the entity linking phase will *automatically* generate a different subset of domain-specific features such as entity_birthPlace, entity_Spouse etc. Based on the celebrity domain training data, the Rank-SVM model may assign higher weights to a different set of features than the ones in movie domain. In this chapter, we present results on *Movie Celebrity* domain which clearly support our claim of WTM being domain independent approach.

## 8.1 Dataset Details

In this section, we have covered the details of preparing the train and test dataset for Celebrities domain.

### 8.1.1   Source of Trivia

Similar to movies, IMDB also provides trivia for Movie Celebrities domain but without
any vote by reader. Figure 8.1 shows a snapshot of Trivia page provided for the celebrity
*Christopher Nolan*. Note that, unlike movie trivia, there are no votes along with Trivia.

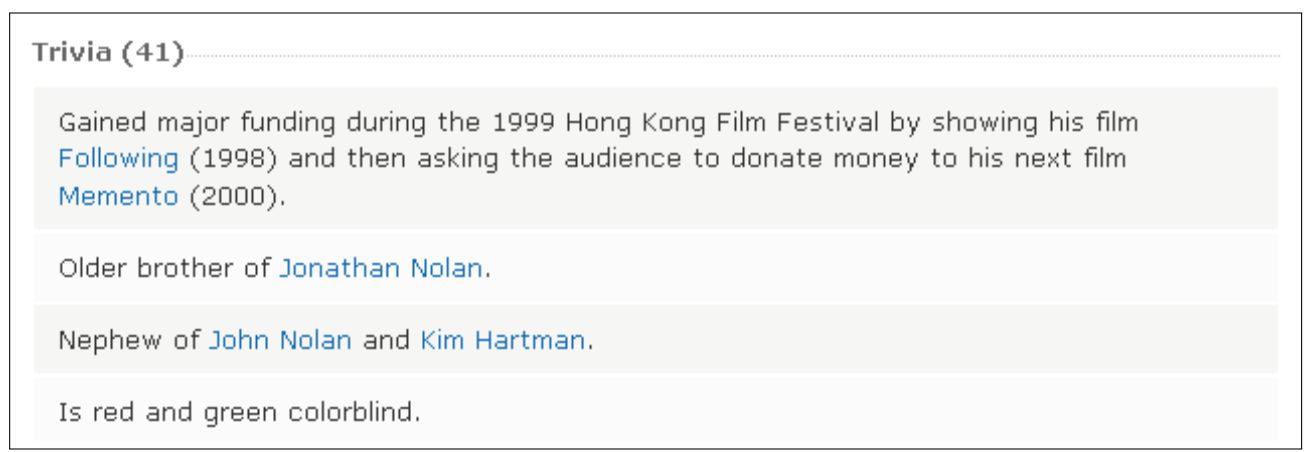

Figure 8.1: Screenshot for Trivia page of *Christopher Nolan* [*ref.* IMDB]

### 8.1.2   Preparation of Dataset

We scrapped all the available trivia for top 1000[1] celebrities, from IMDb. As stated in
Section 8.1.1 the trivia samples are not accompanied by voting data. Hence, we need
to get them judged through crowd-sourcing. Total number of trivia crawled summed to
28158 rows. Getting so many rows judged, is neither economical nor required. So, we
devised a mechanism to prepare a good training set but with reduced size as described
in next paragraph.

Since very long trivia would be time consuming for reading, and in general repelling
for any reader, we set a maximum threshold of 140 characters. Also, since we are getting
the trivia judged from general crowd, we included trivia of popular celebrities so that
there is a high chance that the judges would be knowing the celebrity. For estimating
the popularity of an entity, we used *Google Freebase API*[2] to get search popularity
score. We selected the top 5K rows after sorting the database high to low, according
to celebrity's popularity. Furthermore, since we use pairwise ranking algorithm which
requires comparison of trivia within single celebrity, we selected only those celebrities

---

[1]Listoftop-1000Celebrities:~http://www.imdb.com/list/ls052283250/
[2]https://developers.google.com/freebase/. The API will be retired on June 30, 2015

(and their trivia), which had more than 10 trivia. Finally we obtained 4959 trivia, from 116 celebrities.

We isolated 10 celebrities and their trivia to form test dataset. Rest of the entities and their trivia formed our train dataset. Due to train and test set coming from same source, ranking performance will be higher and lead to higher metrics as compared to the ones reported on movie domain. But when all approaches are applied to the same test set, it is still fair to compare among them. Table 8.1 gives details of train and test dataset prepared.

Table 8.1: Dataset Details (Celebrity Domain)

| Dataset Name | Source | No. of Sent. | No. of Celebri-ties | No. of Pos. Sent. (Trivia) |
|---|---|---|---|---|
| Train Set | IMDB | 4459 | 106 | - |
| Test Set | IMDB | 500 | 10 | 263 |

## 8.2   Features Generated through Entity Linking

In this section, we give examples of entity features generated in celebrity domain. Recall that these features have been generated using `attribute:value` pairs obtained for the entity from DBpedia, as described in Section 4.3.3. Table 8.2 shows some examples of entity features generated along with the sentence, for celebrity domain.

Table 8.2: Sample Entity Features for Celebrity Domain

| | EXAMPLES | |
|---|---|---|
| FEATURE | ENTITY | TRIVIA |
| entity_partner | Johnny Depp | Engaged to _Amber Heard_ [January 17, 2014]. |
| entity_citizenship | Nicole Kidman | First _Australian_ actress to win the Best Actress Academy Award. |
| entity_birthplace | Robert Downey Jr. | Shooting "Sherlock Holmes" in Brooklyn, _New York_.[Jan 2009] |
| entity_alternativenames | Angelina Jolie | Was born Angelina Jolie _Voight_, however in 2002, she dropped her surname "_Voight_" and began using her middle name "Jolie". |

## 8.3  Ranking and Results

We trained $SVM^{rank}$ from the created train dataset, with only two grades - interesting and boring. Reason for taking only two grades, instead of 5 as in movie domain, is less number of judges (votes) for each sample. Due to less votes, Likeness Ratio would be highly discrete. Ranker can be trained with two grades also, but with more grades the ranking quality can be enhanced.

### 8.3.1  Feature Contribution

Table 8.3 presents most weighted features obtained for celebrity domain. Note that the features are intuitive for humans also e.g., word *win*, *nominate*, *magazine* etc. Since the data is from IMDB, many of the sentences would have been about the celebrity being nominated or winning some award. This shows how well the model has been able to generalize and identify which 'about-nesses' are found interesting by majority of people.

Table 8.3: Most informative features (by weight) from Interestingness Ranker Model along with their Feature Group

| Rank | Feature | Group | Weight |
|------|---------|-------|--------|
| 1 | win | Unigram | 0.870133 |
| 2 | nominate | Unigram | 0.867804 |
| 3 | magazine | Unigram | 0.867646 |
| 4 | superPOS | Linguistic | 0.866699 |
| 5 | MONEY | Entity (NER) | 0.861492 |
| 6 | entity_alternativenames | Entity | 0.860835 |
| 7 | root_engage | Linguistic | 0.860835 |
| 8 | entity_citizenship | Entity | 0.859821 |
| 9 | root_die | Linguistic | 0.859819 |
| 10 | foreign | Unigram | 0.854721 |
| 11 | entity_occupation | Entity | 0.837472 |
| 12 | subj_entity_birthname | Linguistic + Entity | 0.837084 |
| 13 | underroot_entity_partner | Linguistic + Entity | 0.834138 |
| 14 | subj_earnings | Linguistic | 0.833371 |
| 15 | subj_entity_children | Linguistic + Entity | 0.823995 |
| 16 | travel | Unigram | 0.817773 |
| 17 | subj_entity_parents | Linguistic + Entity | 0.816521 |
| 18 | entity_birthplace | Entity | 0.808754 |
| 19 | subj_unlinked_location | Linguistic + Entity | 0.798542 |
| 20 | root_love | Linguistic | 0.789954 |

### 8.3.2 Quantitative Comparison

We compared variants of Wikipedia Trivia Miner – WTM (U+L+E) and WTM (U) with
the defined baselines described in Section 6.4. Figure 8.2 shows the actual P@10 obtained.
Note that WTM (U+L+E) has outperformed all other approaches. Increase in P@10 for
WTM (U+L+E) as compared to WTM (U) shows the significance of our sophisticated
language analysis and entity features.

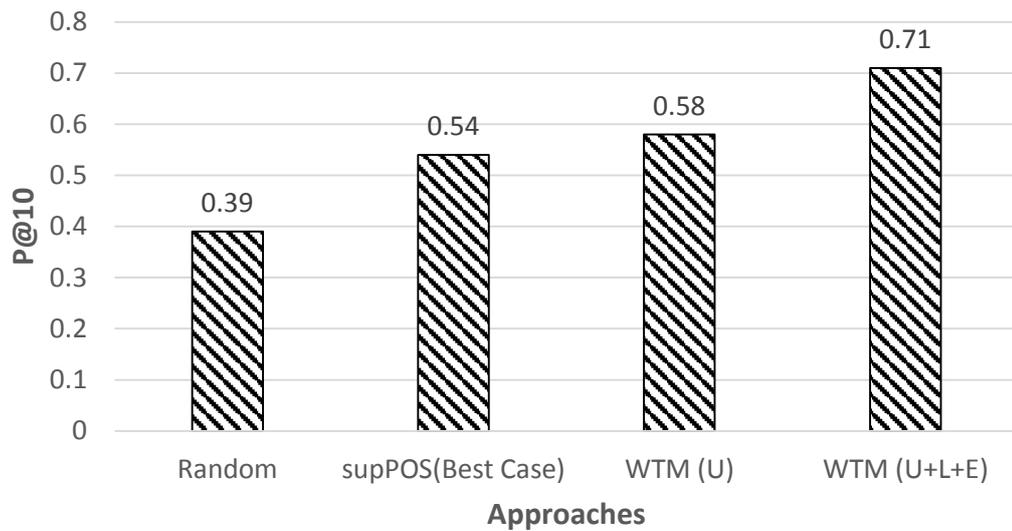

Figure 8.2: P@10 for various approaches (Celebrity Domain)

## Conclusion and Future Work

WE introduced the problem of automatically mining interesting trivia for entities from the unstructured text present at entity's Wikipedia page. We proposed a novel system called "Wikipedia Trivia Miner (WTM)" for tackling the problem and demonstrated its performance on movie entities. Experiments on movie entities and Wikipedia dataset reveal that the proposed system performs significantly better than the baseline approaches and indeed succeeds in surfacing interesting trivia sentences for the entity in focus. WTM approach is domain independent which means for any domain only the train data has to be replaced, which has been demonstrated in Chapter 8. The success of the system could be mainly attributed to the sophisticated domain-independent ranking features which are based on language analysis and entity understanding.

## 9.1 Future Works

Since we are the first one to introduce the problem of mining interesting trivia from text, problem has a lot of scope for future work in multiple directions as discussed in following subsections.

### 9.1.1 Increasing Ranking Quality (P@10) by New Features

In this thesis work, the maximum Precision at 10 (P@10) that we could attain is 0.455, which gives ample room and score of research to increase this value. For this, better set of features can be engineered. One such feature thought was to identify how unusualness of the fact. This can be captured by checking the probability of candidate sentence to occur in the entity's domain. If the probability is very less then it means the sentence states some unusual fact. A good research will be involved to estimate the probability.

Another way to capture unusualness could to detect deviation from expectation in stated fact. This could be used initially may be with the sentences containing numerical figures e.g., revenue of movies. If the revenue deviates a lot from expectation, then it could be treated as trivia.

A trivia would be more surprising and amusing (and hence interesting) if it was not already known to the reader. To estimate the general popularity of the fact stated in the candidate sentence and its effect on ranking quality will be a challenging research problem.

### 9.1.2   Increasing Ranking Quality (P@10) by Other Approach

Another proposal could be to get massive amount of data, and apply Deep Learning to solve the problem. Deep Learning has shown significant results in problem of sarcasm detection, and may show a similar result for trivia detection too.

### 9.1.3   Trying with Classification instead of Ranking

We have tried to solve the problem with Ranking approach, as our scenario required to mine top-k interesting trivia. But researchers can re-formulate the problem for classification task. The application will be highlighting interesting trivia all across the document e.g. on entity's Wikipedia page.

### 9.1.4   Increasing WTM's Recall by replacing references

As discussed in Section 6.1.1, approach for Candidate Selection can be improved to obtain a higher recall. Use of Co-reference resolution could be studied for converting out-of-context sentences to standalone self-sustaining candidates.

### 9.1.5 Increasing WTM's Recall by identifying complete context sections

Quite often, a complete context story is presented in 3-4 adjoining sentences. Such short stories could also provide interesting trivia. A research could be done to identify the collection of minimum number of sentences which present a complete context. Such collections can be taken as candidate for further ranking. They will form multiple sentence trivia, and hence increase the total recall.

### 9.1.6 Personalized Trivia

A fact that is interesting for Reader 1 may not be interesting for Reader 2. A good challenging research can be done to *recommend* trivia according to reader's interest and estimated domain knowledge. This is similar to the task of providing personalized search results for same search query. The personalization can be done on a lot many factors like geographical location, age and other demographics of the reader.

### 9.1.7 Generating Questions from the Mined Trivia

As was stated in Introduction, trivia can be presented in question form also. Although there has been some work done in generating trivia questions from structured databases, generating trivia question from unstructured text will be a new field. The work presented in this thesis can be used to mine interesting trivia in sentence form and then question will have to be generated using that sentence.

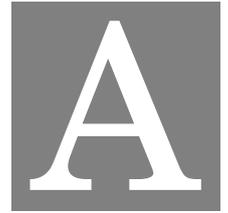

## Actual Results obtained through WTM

Table A.1: Most Interesting (Top-5 scored) Trivia.

| MOVIE | TRIVIA | ACTUAL |
|---|---|---|
| Final Destination (2000) | All death scenes were filmed using lifecasts of the actual actors. | 1 |
| Her (2013) | *Her* also won Best Film and Best Director for Jonze at the National Board of Review Awards, and the American Film Institute included the film in its list of the top ten films of 2013. | 1 |
| Pineapple Express (2008) | *Pineapple Express* received generally positive reviews from critics. | 0 |
| Final Destination (2000) | It received the Saturn Award for Best Horror Film and Best Performance by a Younger Actor for Sawa's performance. | 1 |
| Who Framed Roger Rabbit (1988) | Walt Disney Productions purchased the film rights to the story in 1981. | 1 |

Table A.2: Least Interesting (Bottom-5 scored) Trivia.

| MOVIE | TRIVIA | ACTUAL |
|---|---|---|
| Interstellar (2014) | The film went number one in South Korea ($14.4 million), Russia ($8.9 million) and France ($5.3 million). | 0 |
| Transformers Dark of the Moon (2011) | In China, its highest-grossing market after North America, the film set records for an opening day with $15.9 million, a single day with $17.4 million (overtaken by Journey to the West: Conquering the Demons) and an opening weekend with $46.8 million ($62.7 million with previews). | 0 |
| Man of Steel (2013) | Its opening weekend gross of $116.6 million was the third-highest of 2013, behind Iron Man 3 ($174.1 million) and The Hunger Games: Catching Fire ($158.1 million), and the third-highest among non-sequels, behind Marvel's The Avengers ($207.4 million) and The Hunger Games ($152.5 million). | 1 |
| Interstellar (2014) | In the United Kingdom the film debuted at number one earning Â£5.37 million ($8.6 million) in its opening weekend which was lower than the openings of The Dark Knight Rises (Â£14.36 million), Gravity (Â£6.24 million) and Inception (Â£5.91 million). | 0 |
| Lone Survivor (2013) | Outside of North America, the film's biggest markets were in Australia, the United Kingdom, Spain, Japan, France, South Korea and Germany; the film grossed approximately $3.5 million in Australia, $3.4 million in the United Kingdom, $2.5 million in Spain, $2.2 million in Japan, $1.5 million in France, $1.2 million in South Korea, and $1 million in Germany. | 0 |

# B

## KAPPA STATISTICS

KAPPA Statistics, as described in [24], is a metric to measure how much agreement lies between two independent observers (or judges). The kappa value for two observers may lie in between -1 to 1, where 1 denotes perfect agreement, 0 means results have mean marked randomly and -1 agreement less than chance, i.e. potential systematic disagreement between observers. Suppose, for two observers samples are marked as Table B.1 (taken from [24]):

Table B.1: Representative Observation Table.

| | | Observer 1 - Result | | |
|---|---|---|---|---|
| | | Yes | No | Total |
| **Observer 2 - Result** | Yes | *a* | *b* | $m_1$ |
| | No | *c* | *d* | $m_0$ |
| | Total | $n_1$ | $n_0$ | $n$ |

From this data, Expected Agreement is calculated by Eqn. B.1

$$p_e = \left[ \left( \frac{n_1}{n} \right) * \left( \frac{m_1}{n} \right) \right] + \left[ \left( \frac{n_0}{n} \right) * \left( \frac{m_0}{n} \right) \right] \tag{B.1}$$

Observed Agreement is calculated by Eqn. B.2

$$p_o = \frac{(a+d)}{n} \tag{B.2}$$

Using the above two calculated values, Kappa is calculated by Eqn. A.3

Observed Agreement is calculated by Eqn. B.3

$$K = \frac{(p_o - p_e)}{(1 - p_e)} \tag{B.3}$$

Based on the obtained Kappa value, the conclusion on agreement can be made according to the given Table B.2 (taken from [24]).

Table B.2: Interpretation of Kappa.

| Kappa | Agreement |
|---|---|
| < 0 | Less than chance agreement |
| 0.01-0.20 | Slight agreement |
| 0.21-0.40 | Fair agreement |
| 0.41-0.60 | Moderate agreement |
| 0.61-0.80 | Substantial agreement |
| 0.81-0.99 | Almost perfect agreement |

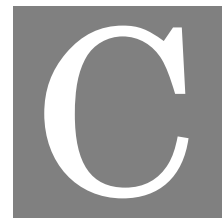

# RANKING V/S CLASSIFICATION

SINCE our problem focuses on retrieving the top-k most interesting trivia, ranking was the natural choice over classification. Still, in this section we have reported metrics obtained using Classification approach too. Also, we have discussed the reason for ranking outperforming classification for the problem and the evaluation metrics we have formulated.

In Classification approach, we trained a SVM model with the grade/class as ground label, and obtained parallel class/grade value for the unseen Wikipedia candidates. We sorted results by the descending order of obtained grade label, and for the ones with the same label, we sorted them in decreasing value of probability of being in that label, scored by the classifier. As described in Section 6.3, metric P@10 is reported by using only two scale (1/0) class value only.

Table C.1: Metrics Comparison of Classification and Ranking.

| Method | P@10 |
|---|---|
| Classification (1/0 grades) | 0.31 |
| Ranking (1/0 grades) | 0.38 |
| Classification (4-0 grades) | 0.34 |
| Ranking (4-0 grades) | 0.455 |

Table C.1 shows the different possibilities with Ranking and Classification approaches. Note that Ranking with 5 grades has outperformed all other approaches.

SVM Ranking approach outperforms Classification over the metrics specific to ranking, because of the Cost function it uses as the feedback of error. Figure C.1, taken from [25], shows how different (wrong) ordering has the same value of classification cost function, whereas has different (more accurate) values for ranking cost function. Ranking cost function gives a true feedback of wrong ordering, and hence more effective.

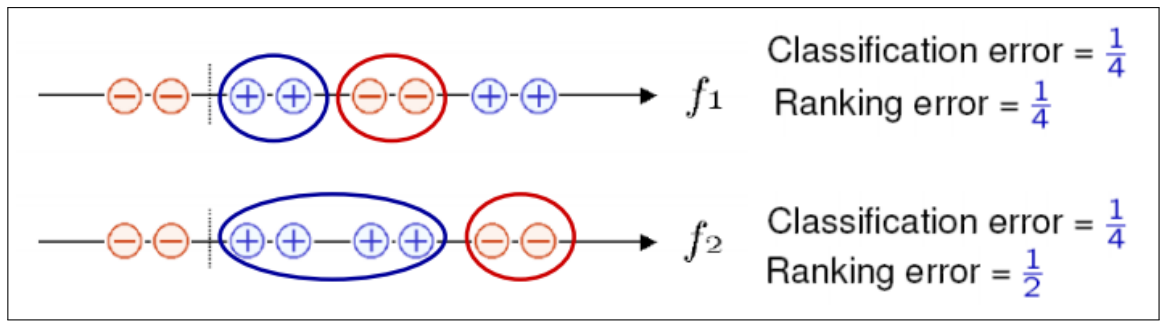

Figure C.1: Cost Function in Classification v/s Ranking

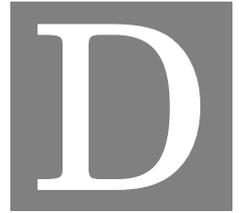

APPENDIX D

# SAMPLE OF ACTUAL ENTITY TRIVIA FOUND IN WIKIPEDIA

Table D.1: Sample Trivia from Wikipedia

| ENTITY | TRIVIA |
|---|---|
| Bill Gates | He was arrested in New Mexico in 1977 |
| | For a brief period in 1999, his net worth went past $100 billion |
| | In the first five years, Gates personally reviewed every line of code the company shipped |
| | He scored 1590 out of 1600 on the SAT |
| | His first office for Micro-Soft was in Albuquerque |
| Amitabh Bachchan | He started his film debut as a voice narrator |
| | He is the most-nominated performer in any major acting category at Filmfare, with 39 nominations overall |
| | His father was Hindi poet |
| | His earlier name was Inquilaab |
| | His name 'Amitabh' was a suggestion of Sumitranandan Pant |
| | His actual surname was Shrivastava |

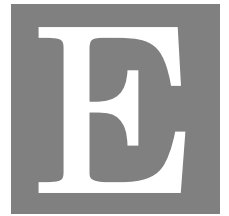

# RESOURCES: DATA, CODE AND DEPENDENCIES

We have shared all the resources – Data and Code publicly. The project has few external dependencies too. The locations to avail the resources have been given below:

1. **Data**: `https://github.com/abhayprakash/WikipediaTriviaMiner_SharedResources/tree/master/1_Data`

2. **Code**: `https://github.com/abhayprakash/WikipediaTriviaMiner_SharedResources/tree/master/2_Code`

3. **Documentation and Binaries of** $SVM^{rank}$: `http://svmlight.joachims.org/`

4. **Wikipedia & IMDB Crawlers**: `https://github.com/abhayprakash/triviaGeneration/tree/master/TriviaGeneration/TriviaGeneration`

5. **Github Repository with version history**:
`https://github.com/abhayprakash/triviaGeneration/`